\documentclass{aa}
\newcommand\x{0.25}
\newcommand\diff{1.5\times10^{-2}}
\newcommand\pr{0.06}
\newcommand\pmag{4}

\newcommand\Nbin{340}
\newcommand\Nphbin{4{\times}10^4}

\newcommand\tend{100}
\newcommand\ani{2.5}
\newcommand\li{315}
\newcommand\licustom{84}

\newcommand\rhoo{1.6{\times}10^{-27}}
\newcommand\To{6.2}
\newcommand\texp{2}
\newcommand\ksisim{28}
\newcommand\elliL{22}
\newcommand{\zstroke}{%
  \text{\ooalign{\hidewidth -\kern-.3em-\hidewidth\cr$z$\cr}}%
}
\newcommand\Hrho{0.3}
\newcommand\HT{2.8}
\usepackage{natbib}
\bibpunct{(}{)}{;}{a}{}{,}

\defcitealias{perrone2D}{PL22a,b}
\defcitealias{perrone3D}{PL22b}

\usepackage{graphicx}
\usepackage{txfonts}
\usepackage[breaklinks, colorlinks, citecolor=blue, urlcolor = blue]{hyperref}
\begin{document} 

   \title{Dynamical properties and detectability of the magneto-thermal instability in the intracluster medium}

   \author{J. M. Kempf
          \inst{1},
          F. Rincon
          \inst{1}
          \and
          N. Clerc \inst{1}
          }

   \institute{Institut de Recherche en Astrophysique et Planétologie (IRAP), Université de Toulouse, CNRS, UPS, Toulouse, France\\
              \email{jean.kempf@irap.omp.eu}
             }

   \date{\today}

 
  \abstract
   {
   Amongst the many plasma processes potentially relevant to the dynamics of the intracluster medium (ICM),
   turbulence driven at observable scales by
   internal magnetised buoyancy instabilities such as the magneto-thermal instability (MTI)
   stands out in the outskirts of the ICM,
   where the background temperature decreases with the radius.
   }
   {
   We characterise the statistical properties of MTI turbulence in the ICM 
   and assess whether such large-scale magnetised plasma dynamics would be detectable
   with the future X-ray calorimeter X-IFU on board ATHENA.
   }
   {
   We made use of scaling laws previously derived
   to phenomenologically estimate the observable turbulent saturation levels
   and injection length of MTI turbulence
   for different ICM thermodynamic profiles,
   and performed a numerical magnetohydrodynamic simulation
   of the dynamics
   with Braginskii heat and momentum diffusion.
   As a prospective exercise,
   we used the simulation to virtually
   observe MTI turbulence
   through the X-IFU.
   }
   {
   In bright enough regions amenable to X-ray observations, the MTI drives mild turbulence
   up to $\sim 5\%$ and $\sim 100$ km/s  (root-mean square temperature fluctuation and velocity).
   However, the measurable integrated temperature fluctuation
   and line-of-sight velocity fields,
   the latter being essentially the azimuthal velocity component in cluster haloes,
   hardly exceed $1\%$ and $10$ km/s, respectively (root-mean square).
   We show that such moderate signals would be difficult to detect
   with upcoming X-ray telescopes.
   We also find that
   MTI turbulence 
   is anisotropic in the direction of gravity
   and develops at scales
   ${\gtrsim}0.2$ Mpc.
   If the fluctuation intensities were to be stronger
   than the current theoretical estimates,
   MTI fluctuations would be detectable
   and their anisotropy
   discernible with the X-IFU.
   }
   {
   Finding direct signatures of magnetised plasma dynamics in the ICM,
   even at observable scales typical of the fluid MTI, remains challenging.
   This study only marks a first step in this direction.
   Several numerical and observational strategies are discussed
   to make further progress in the future.
   }


   \titlerunning{Dynamical properties and detectability of the MTI in the ICM}
   \maketitle
%

\section{Introduction}

   Galaxy clusters
   are filled with hot and diffuse gas:
   $k_\mathrm{B} T{\sim}5\ \mathrm{keV}, \ \rho{\sim}10^{-27}\ \mathrm{g/cm}^{3}$.
   This plasma, usually referred to as the intracluster medium (ICM),
   radiates in the X-ray
   through the combined emission of lines and Bremsstrahlung
   at a very high temperature.
   It accounts for ${\sim} 15 \%$
   of the total mass of clusters and
   is in rough hydrostatic equilibrium
   within its dark matter potential well.
   Recent observations however
   suggest that the ICM sustains subsonic turbulence
   \citep{zhuravleva2018,dupourque},
   which could alter the hydrostatic balance
   and the internal transport
   of chemicals and energy,
   and hence shaping the evolution of clusters
   on cosmological timescales \citep{zhuravleva}.
   A departure from ideal hydrostatic equilibrium in the ICM could also
   provide additional non-thermal pressure support \citep{eckert},
   biasing the determination of cluster masses
   through X-ray and Sunyaev-Zeldovich (SZ) measurements
   \citep{vazza,angelinelli}.
   The assessment of such a systematic bias is critical for cosmology
   since the mass distribution of galaxy clusters is used
   as a late-universe probe for the cosmological parameters
   (see the reviews by \citealt{allen,pratt}).
   
   The dynamics of the ICM
   remains observationally poorly constrained.
   The Soft X-ray Spectrometer (SXS)
   calorimeter on board the Hitomi satellite made
   the only direct measurement of a velocity field in the ICM
   of the Perseus cluster core
   \citep{hitomi16, hitomi18}.
   The instrument was able to measure velocity gradients
   and turbulent broadening about $150$ km/s
   thanks to both an emission line centroid shift and width measurements, respectively,
   but only down to the SXS pixel resolution of $20$ kpc.
   So far, putting observational constraints on ICM turbulence below this
   length scale is exclusively achievable
   with the detection of ICM thermodynamic perturbations \citep{hofmann}
   or thanks to
   indirect measurements of its kinematics
   \citep{simionescu2019}
   through measurements of
   X-ray surface brightness
   \citep{churazov2012, zhuravleva2015}
   and SZ
   fluctuations
   \citep{sunyaev, mroczkowski}.
   However, the next generation of spatially resolved
   X-ray spectroscopy instruments, such as Resolve \citep{xrism}
   or the X-ray Integral Field Unit \citep[X-IFU;][]{barret16,barret23}
   on board the future X-ray observatory ATHENA, and their unprecedented spectral resolution
   ($7$ and $2.5$ eV below $7$ keV, respectively)
   will push this limit downwards and
   provide us with an unprecedented opportunity to characterise
   the ICM dynamics
   down to the kiloparsec scale for the closest clusters.

   The possible sources of turbulence in galaxy clusters are numerous
   and diversely distributed across the ICM,
   ranging from active galactic nuclei (AGN) feedback at the centre
   \citep{conroy,gitti}
   to accretion of infalling baryons on the outer edge
   \citep{vazza2017, iapichino}
   and ram-pressure stripping in between
   \citep{domainko, ebeling, li} for instance.
   Internal fluid instabilities may also be important in this context.
   The ICM is
   stably stratified
   against thermal convection
   \citep{cavagnolo, ghirardini}
   according to the usual
   Schwarzschild criterion on the entropy gradient
   $\partial_r \log \left( T \rho^{1-\gamma} \right)>0$ \citep{schwarzschild},
   with $\gamma$ being the adiabatic index.
   However, it is magnetised up to
   $B {\sim} 1{-}10 \ \mathrm{\mu G}$
   \citep{govoni,ferrari,botteon22}.
   ICM plasma
   is therefore thought to be in a 'dilute' regime
   where the particles' mean-free path is much larger
   (a dozen orders of magnitude)
   than the particles' Larmor radius.
   This ordering introduces an
   anisotropy
   with respect to the direction
   of the local magnetic field:
   heat and momentum collisionally diffuse primarily along the magnetic
   field lines and barely across them \citep{braginskii}.
   In these conditions, \citet{balbus2000,balbus2001}
   showed that the sign of the temperature (and no longer the entropy) gradient dictates
   the stability of the stratified fluid.
   In particular, a so-called magneto-thermal instability (MTI)
   is triggered
   when the background temperature gradient is in the direction of the gravity,
   and may excite turbulence.
   This magnetised buoyancy instability is to dilute plasma
   what classical thermal convection is to usual gas.
   As seen from XMM-Newton
   \citep{leccardi}
   and Chandra
   \citep{simionescu}
   observations (sometimes combined for a SZ analysis,
   \citealt{shitanishi,ghirardini}),
   all galaxy clusters exhibit a
   decreasing temperature profile
   (though as of a certain radius for relaxed clusters)
   and should therefore be unstable to the MTI, at least in the halo.

   Magnetised buoyancy instabilities such as the MTI
   and its counterpart, the Heat-flux driven Buoyancy Instability
   (HBI; triggered when the temperature increases with radius, \citealt{hbi}),
   have been theorised two decades ago and their
   potential relevance in the ICM has been largely
   emphasised since then
   \citep{bogdanovic, balbus2010, mccourt12, parrish2012a, kunz, kunzlatter, berlok16a}.
   For example, they
   are thought to drive a substantial magnetic field amplification through a dynamo,
   and thus making them a serious candidate
   mechanism for cluster magnetisation.
   Magnetic seeds in galaxy clusters
   (e.g. stemming from primordial magnetic fields, \citealt{durrer})
   would indeed need to grow by several orders of magnitude
   to match the strength currently observed in neighbour clusters,
   although strong enough primordial magnetic seeds
   could also lead up to the present magnetic strengths
   through compression only
   (\citealt{jedamzik}; see \citealt{donnert}
   for a review on magnetic field amplification in galaxy clusters).
   Such magnetised buoyancy instabilities
   remain however somewhat
   speculative since no observational evidence
   of their presence has ever been detected.
   An additional complexity stems from the conservation of the adiabatic invariant
   $\mu \propto \varv_\perp^2/B$ (with $\varv_\perp$ being the perpendicular component
   of the particle's peculiar velocity with respect to the magnetic field $\vec{B}$),
   which 
   drives pressure anisotropy
   when coupled to changes in magnetic-field strength
   in high-beta and weakly collisional plasma.
   Such pressure anisotropies can, when large enough, trigger mirror, firehose, ion-,
   and electron-cyclotron
   whistler instabilities \citet{schekochihin05,schekochihin10}.
   The saturation of such micro-scale kinetic instabilities
   could severely reduce and/or isotropise the heat conductivity
   \citep{riquelme, komarov}
   and hamper the development of the MTI
   in the ICM,
   if not completely annihilate it \citep{drake};
   although, recent work by \citet{berlok21} tends to show
   only modest consequences of these micro-instabilities
   on the MTI saturation.

   Overall,
   detecting the MTI, if it is present in the ICM and
   if detectable with future X-ray missions, would be a very interesting
   step forward, not only from a purely observational perspective
   to characterise turbulent transport or magnetisation in the ICM,
   but also from a theoretical and astrophysical plasma physics standpoint.
   It would open a direct observational window into the physics of dilute and
   magnetised astrophysical plasmas on large fluid scales,
   where previous work by \citet{zhuravleva2019} could only rely on indirect signatures
   of micro-scale plasma processes
   such as the enhanced collision rate possibly due to
   the previously mentioned kinetic instabilities.
   In that paper, the authors used deep Chandra observations of the Coma cluster
   to probe the ICM dynamics through X-ray surface brightness fluctuations down to
   what should be the viscous scale
   if only Coulomb collisions were responsible of momentum transport.
   They could not detect any effect of the viscosity at this scale
   and therefore deduced that the ICM plasma is either subject to enhanced collision rates
   or to Braginskii anisotropic transport.
   The interest for such new observational windows goes beyond the ICM because
   the same plasma regime
   is also relevant in accretion flows around black holes for instance
   \citep{schekochihin07,sharma}.
   The current study is therefore intended to be a first step towards
   bridging the gap between 
   theoretical astrophysical plasma studies and future X-ray astronomy observations.

   The outline of this paper is as follows.
   In Section \ref{sec:method}, we introduce
   different thermodynamic profile models of the ICM,
   and we describe
   the magnetohydrodynamics (MHD) model and the numerical methods used to simulate
   MTI turbulence,
   as well as the post-processing pipeline developed
   to build synthetic observations of the MTI through the X-IFU instrument.
   We use the X-IFU specifications as defined before the recent design-to-cost exercise \citep{barret16,barret23}
   as a potential model of a future X-ray spectrometer:
   it acts as our baseline instrumental configuration throughout this paper.
   In Section \ref{sec:char}, we characterise
   the phenomenological dynamical properties
   of MTI-driven turbulence expected in the ICM,
   relying on scaling laws derived recently
   by \citet[][respectively \citetalias{perrone2D} hereafter]{perrone2D, perrone3D}
   and on the ICM thermodynamic models previously introduced.
   The objective of that section is to provide
   a first estimate
   of the expected levels of measurable MTI turbulence in the ICM,
   taking into account various observational effects such as
   the finite value of the ATHENA/X-IFU effective collecting area
   or the cancellation of opposite-sign fluctuations present along the line of sight.
   In Section \ref{sec:results},
   we present the numerical results from our Braginskii-MHD simulation
   and discuss the consequences of the discrepancy
   between numerical and real ICM plasma regimes,
   and the need for a rescaling of the simulation to perform mock observations.
   As a prospective exercise that may also prove of interest to theoreticians to fix ideas
   about the observability of the MTI,
   we then present two synthetic observations of the MHD simulation scaled in different ways.
   We qualitatively compare
   the output velocity
   and thermodynamic fluctuation fields
   that can be reconstructed from these observations
   with the raw input fields from the rescaled simulation.
   Using one of the two synthetic observations,
   we observationally constrain the anisotropic nature of the MTI.
   Our main conclusions
   are summarised
   in Section \ref{sec:discussion},
   where we also stress the caveats or the current work
   along with its implications
   for future X-IFU observations.
   We finally extend the discussion to other sources of anisotropic turbulence in the ICM.

\section{Methods and models}
\label{sec:method}
   In this section, we first present three different thermodynamic profile models of the ICM,
   which will be used
   in Section \ref{sec:char}
   to phenomenologically discuss
   the properties of MTI-driven turbulence in clusters.
   Based on the expected number of photons
   from a 2-Ms observation with the X-IFU
   (which we use as a prospective model
   of X-ray spectrometer),
   we also give a criterion
   to determine an admissible range of radii at which measurement uncertainties
   due to a lack of photons are expected
   not to be predominant over the instrumental limitations themselves.
   We then describe the physical MHD model
   used to numerically study the development
   of the MTI up to a state of sustained turbulence.
   Finally, we present the methods used to build synthetic X-IFU observations of a MHD simulation.

   Throughout this work, we assume a universe with a
   $\Lambda$-CDM cosmology and
   $H_0=70 \ \rm{km/s/Mpc}$,
   $\Omega_\mathrm{m}=0.3$, $\Omega_\Lambda=0.7$.
   We set the primordial helium and metal mass abundances
   to $Y=0.24$ and $Z=0.06$ respectively.
   The virial radius $R_{200}$ is defined as the radius within which
   the average matter density is $200$ times higher that the
   critical density of the universe
   $\rho_c= 3H^2/(8\pi G)$
   where $H$ is the Hubble parameter and $G$ the universal gravitational constant.
   Although this radius varies from a cluster to another,
   we set it to a standard value of $R_{200}{=}1.8$ Mpc
   for all clusters.
   The symbol $\zstroke_0$ (resp. $\zstroke$)
   denotes the cosmological (resp. cosmological plus Doppler) redshift
   whereas the letter $z$
   always represents a vertical coordinate in Cartesian geometry
   (in spherical geometry, $r$ is the radial coordinate).
   $\vec{\varv}$ and $\vec{B}$ respectively stand for the velocity and magnetic field.
   The unit vector in the direction
   of the local magnetic field is $\vec{\hat{b}}=\vec{B}/B$.
   The pressure, density, temperature and entropy are respectively denoted
   by $p$, $\rho$, $T$ and $S=T\rho^{1-\gamma}$
   (the adiabatic index $\gamma$ equals $5/3$ for mono-atomic perfect gas).
   Their respective fluctuations are always preceded by the lower-case Greek $\delta$
   and expressed in a non-dimensional way:
   $\delta X = \left(X-\overline{X}\right)/\overline{X}$ where $X$ is one of the previous
   thermodynamic quantity and $\overline{X}$ its average at iso-gravity.
   The letter $\vec{g}$ stands for the gravity
   and $g_0$ for its intensity. 
   The thermal diffusivity, the kinematic viscosity and the magnetic resistivity are represented by
   $\chi$, $\nu$ and $\eta$ respectively.
   The scale-height of a physical quantity $X$
   is $H_X\equiv\left(\partial_r \log X\right)^{-1}$.
   The symbol $<\cdot>$ denotes a spatial average, while the root-mean square
   of the physical quantity $X$ is defined as $\left. X \right|_{\mathrm{rms}}^2 = <X^2>$
   since we systematically assume $<X>=0$ for turbulent fluctuations.

   \begin{figure}
   \centering
   \includegraphics[width=0.9\hsize]{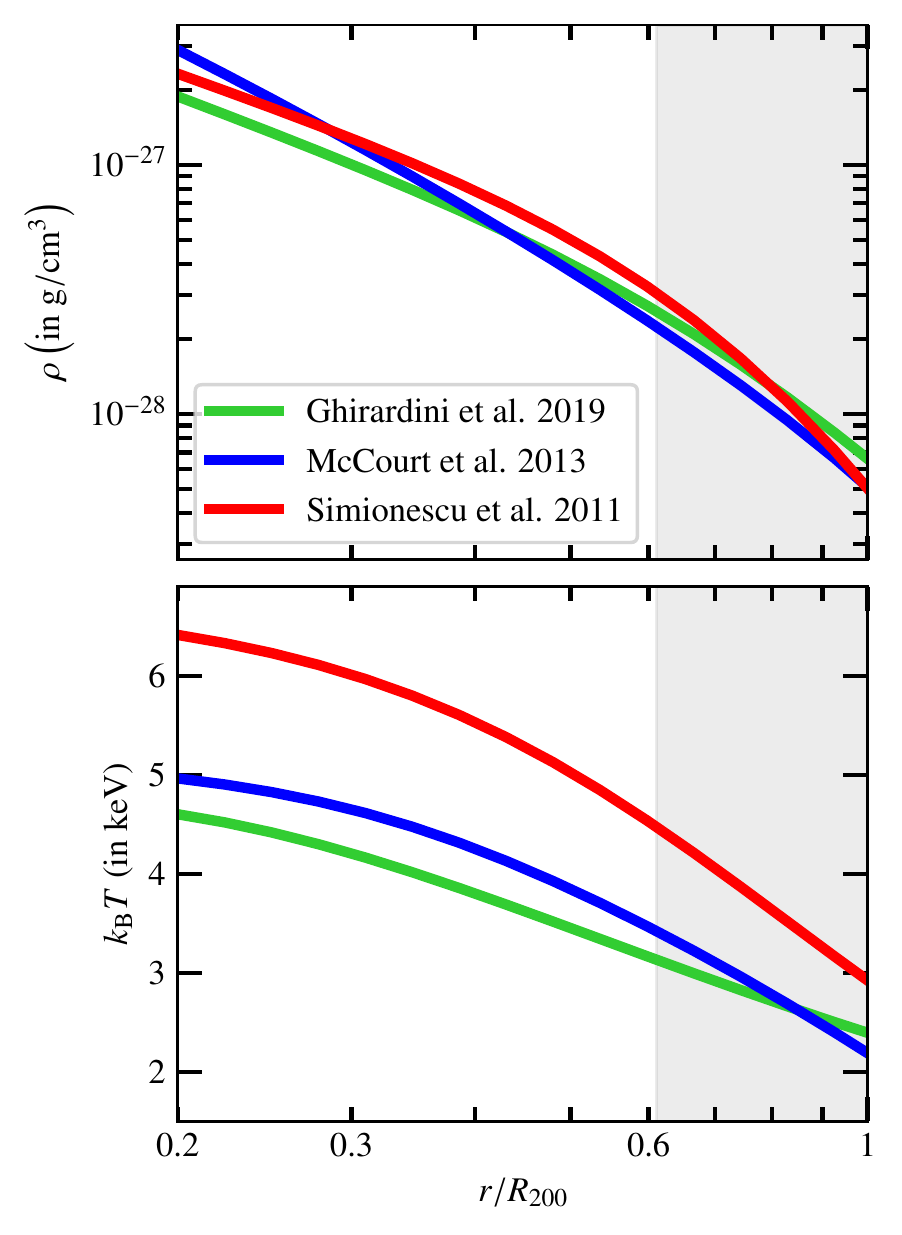}
      \caption{Density $\rho$ (top) and temperature $T$ (bottom)
               profiles as a function of radius
               for three typical ICM models
               \citep{ghirardini, mccourt, simionescu}.
               Only the range with decreasing temperature is shown,
               until $R_{200}$.
               The grey shaded areas highlight radii for which
               the X-IFU will collect less than
               two million photons
               in the 0.2-12 keV energy range
               during a 2-Ms observation.
              }
         \label{fig:profiles}
   \end{figure}

   \subsection{Thermodynamic models of ICM}
   \label{sec:prof}
   We introduce
   three different density and temperature
   profiles that will be later used in Section \ref{sec:char} to show
   that the observable characteristics of
   MTI-driven turbulence
   in galaxy clusters
   are robust against 
   the choice of thermodynamic model.
   The first density and temperature profiles (in green in Fig. \ref{fig:profiles}) comes
   from \citet{ghirardini},
   in which universal thermodynamic profiles are presented for typical cool core (CC)
   and non-cool core (NCC) galaxy clusters as deduced 
   from a joint X-ray and SZ analysis of a X-COP sample with 12 different clusters.
   More specifically, we used the functional forms from
   Sections 3.2 and 3.4 of their paper along with the best-fit
   parameters given in their Table 3 for CC galaxy clusters.
   The reason behind this choice
   is that dynamically relaxed CC clusters
   are a priori better suited to detect MTI turbulence
   than their NCC counterparts whose dynamics at Mpc scales
   is certainly dictated, or at least polluted, by merger events
   rather than MHD instabilities. 
   The second profiles (in blue in Fig. \ref{fig:profiles}) are computed from a
   1D-spherical model by \citet{mccourt},
   which we have independently reproduced.
   This model assumes hydrostatic equilibrium of the plasma
   in a Navarro-Frenk-White gravitational dark matter potential \citep{nfw}
   for a type III accretion history
   \citep{mcbride}
   with a concentration parameter of 5 and no
   effect of thermal conduction.
   Finally, as our closest cosmic neighbour, current X-ray missions have dedicated
   large exposure times to very deep observations of
   the Perseus cluster \citep{fabian2003,fabian2006},
   which is a relaxed CC cluster \citep{simionescu}.
   Perseus-like thermodynamic profiles (in red in Fig. \ref{fig:profiles})
   are therefore well suited for this study
   and we chose them to complete our set of ICM profiles. In practice, we extracted the
   data corrected for clumping of the Perseus north-west arm,
   available in \citet{simionescu}, and we fitted them with the same
   functionals as those used in \citet{ghirardini}.
   The detailed form of the three thermodynamic profiles
   can be found in Appendix \ref{app:prof}.
   For the three models, only radii between $0.2R_{200}$ and $R_{200}$
   are considered,
   as this range ensures that all temperature profiles are decreasing with radius,
   thus meeting the primary criterion for the MTI to be triggered.
   The chosen outer radius is equal to the largest radius at which thermodynamic data
   were available from observation in \cite{simionescu, ghirardini}.

   All regions with decreasing temperature
   are not necessarily bright enough to make their observation in X-ray possible
   in a reasonable amount of exposure time,
   even with the large effective collecting area of the ATHENA/X-IFU \citep{barret16}.
   The plasma emissivity is indeed proportional to the squared density
   (but also non-trivially depends on the temperature
   and metallicity; \citealt{rybicki}).
   Assuming uniform temperature and metallicity fields,
   the knowledge of the density profile
   determines the line-of-sight integrated emissivity
   (i.e. the emission measure Eq. (\ref{eq:EM}))
   and therefore the expected number of photons in a certain range of energy
   for a given exposure time,
   when combined with the X-IFU effective area.
   We choose to virtually observe bright enough cluster regions only,
   which would provide more than $2\times10^{6}$ photons in the $0.2-12$ keV range
   during a 2-Ms X-IFU observation of a Perseus-like cluster.
   In our case, it consists in excluding regions beyond $0.6R_{200}$
   (grey shaded areas in Fig. \ref{fig:profiles}-\ref{fig:lengths}-\ref{fig:3Drms}).
   Below this threshold (i.e. beyond the latter radius),
   we consider X-IFU observation impossible
   because the uncertainties related to the Poisson noise due to a lack of photons
   certainly exceed the instrumental limitations.
   
\subsection{Numerical MHD model}
\label{sec:MHD}
   We now detail the physical numerical MHD model
   used to study the MTI at saturation.
   First, the equations and the numerical methods
   implemented to evolve them are introduced.
   We then describe the initial hydrostatic equilibrium
   along with the numerical parameters
   of our Braginskii-MHD simulation of MTI turbulence.

\subsubsection{Equations and numerical methods}
   The model relies on the usual system
   of compressible MHD equations in conservative forms,
   where stratification and non-ideal effects are accounted for;
   namely magnetic resistivity and anisotropic heat and momentum diffusion.
   In practice, we use the same equations as in \citet{parrish2012b}
   Eqs. (3-9) with an additional magnetic resistive term
   and without the Bremsstrahlung cooling term.
   This system of equations is closed
   thanks to the usual equation of state for a perfect gas.

   As a consequence of the ICM dilute plasma regime,
   the transport of heat and momentum is anisotropic
   with respect to the direction of the local magnetic field.
   The Braginskii heat flux and viscous stress tensor take the
   following forms:
\begin{equation}
   \mathcal{Q} = - n_\mathrm{e} k_\mathrm{B} \chi \vec{\hat{b}\hat{b}} \cdot {\vec{\nabla}} T,
   \label{eq:qbrag}
\end{equation}
\begin{equation}
   \Pi = - 3 \rho \nu \left( \vec{\hat{b}\hat{b}} : \nabla \vec{\varv} - \frac13 \nabla \cdot \vec{\varv} \right) \left(\vec{\hat{b}\hat{b}} - \frac13 \mathrm{I} \right),
   \label{eq:pibrag}
\end{equation}
   where we recall that $\vec{\hat{b}}=\vec{B}/B$ is the unit vector
   in the direction of the magnetic field.

   These Braginskii-MHD partial differential equations are solved in Cartesian geometry using IDEFIX,
   a new finite-volume MHD code for astrophysical fluid dynamics
   \citep{lesur}.
   Amongst the vast choice of available Riemann-solvers, we selected
   HLLD associated with a third-order reconstruction scheme which
   proved to be robust enough.
   The physical quantities are integrated in time using a third-order Runge-Kutta algorithm.
   We developed and included two additional physical modules
   accounting for both the anisotropic heat and
   momentum diffusions.
   Standard tests
   are presented in Appendix \ref{app:brag},
   validating the implementation of the Braginskii operators.
   All parabolic terms can be integrated explicitly at each time step,
   but a Runge-Kutta-Legendre (RKL) super time-stepping scheme
   is also available to speed up this integration.
   We opted for the latter after making sure
   that the results we obtained with both RKL and fully explicit integrations were identical
   for a reduced set of numerical setups.

\subsubsection{Simulation configuration and hydrostatic equilibrium}
\label{sec:config}
   The Braginskii-MHD equations are integrated in a local cubic box of size 
   $L{=}1.5$, with lengths normalised by the sixth of the temperature
   scale-height $H_T$ (i.e. $H_T/6{=}1$ in this setup).
   The box is thus $H_T/4$ long and spans $25\%$ of the temperature scale-height
   with 256 points regularly spaced in each direction.
   Velocities are normalised
   by the thermal velocity
   $\varv_{\mathrm{th},0}{=}\sqrt{T_0}$
   at the bottom of the atmosphere.
   The magnetic energy is normalised as a kinetic energy
   and magnetic fields therefore by $\sqrt{\rho_0}\varv_{\mathrm{th},0}$.
   In these units, times are normalised by $t_0{=}H_T/(6 \varv_{\mathrm{th},0})$.
   The simulation is integrated for $100t_0$
   (roughly $8$ turnover times, see Section \ref{sec:results}).
   The numerical parameters of the setup are the 
   thermal diffusivity
   $\chi{=}\diff$,
   the Prandtl number
   $\mathrm{Pr}{=}\pr$
   and its magnetic counterpart
   $\mathrm{Pm}{=}\pmag$.
   We also define the magnetic Reynolds number
   $\mathrm{Rm} = \left.\varv\right|_\mathrm{rms}/(k_i\eta)$,
   where $\left.\varv\right|_\mathrm{rms}$ and $k_i$ are respectively
   the root-mean square velocity and the integral scale
   that will be determined a posteriori from the simulation.

   In what follows, $(\vec{e_x}, \vec{e_y})$ is the horizontal plane
   and $-\vec{e_z}$ the direction of the gravity $\vec{g} = -g_0 \vec{e_z}$.
   The local model of stratified atmosphere is initialised
   with the hydrostatic equilibrium from \citet{parrish2012b}:
\begin{equation}
   T(z) = T_0 \left(1 - \frac{z}{H_T}\right),
\label{eq:Tz}
\end{equation}
\begin{equation}
   g_0 = \frac{2+\alpha}{H_T}T_0,
\label{eq:g0}
\end{equation}
\begin{equation}
   \rho(z) = \rho_0\left(1-\frac{z}{H_T}\right)^{1+\alpha},
\label{eq:rhoz}
\end{equation}
   where $T_0$ and $\rho_0$ are respectively
   the temperature and density at the bottom of the atmosphere, 
   and $\alpha$ a free non-dimensional parameter controlling the level
   of stratification.
   We set $\alpha{=}2$, $H_T{=}6$, $T_0{=}\rho_0{=}1$.
   All components of the initial velocity field follow
   a random white noise with an amplitude of
   $10^{-4}$.
   Both horizontal components of the
   magnetic field
   are initialised
   with an amplitude
   of $10^{-5}$.
   In the horizontal directions,
   we chose periodic boundary conditions (BC) for all fields
   and, in the vertical direction,
   we implemented a quasi-periodic BC as described
   by \citet{berlok16b}.
   This vertical BC
   allow us to hold both the background hydrostatic equilibria 
   and the background temperature
   profile all along the simulation,
   while periodising the velocity, the magnetic field and the thermodynamic perturbations.
   Thus, we do not need to 'sandwich'
   the MTI-unstable layer between two stable layers with
   isotropic heat conduction, as usually done
   in such MTI configurations
   (\citetalias{perrone3D}; \citealt{mccourt11,parrish05}).

\subsection{Virtual X-IFU observations}
\label{sec:synthetic}
\begin{figure}
   \centering
   \includegraphics[width=\hsize]{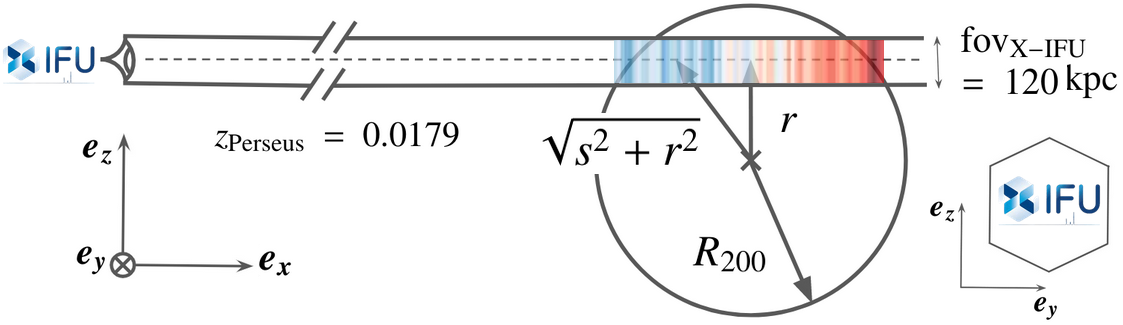}
      \caption{Problem geometry and schematic
               of statistical
               fluctuation cancellation
               for an observation of an ICM at Perseus redshift
               through the X-IFU, which observes the plane of the sky $(\vec{e_y},\vec{e_z})$.
               The coloured boxes can represent either temperature, density or velocity fluctuations.
               The colour scale is arbitrary but blue and red tones are
               of opposite signs, and the more opaque
               the stronger the fluctuation.
               The dependency of the fluctuations with $y,z$
               is overlooked for clarity.
              }
      \label{fig:schema}
   \end{figure}

   We finally sketch how to construct synthetic X-IFU observations
   from our numerical MHD simulation.
   The first step is to extract, from a Braginskii-MHD simulation,
   3D spatial $(x,y,z)$-representations of the density, temperature and velocity fields
   adapted to a mock observation.
   The procedure is quite technical, and 
   thus described in details in Appendix \ref{app:design}.
   The geometrical configuration of the final box
   is always a parallelepiped rectangle,
   with equal lengths in
   the plane of the sky $(\vec{e_y}, \vec{e_z})$
   but elongated along the $\vec{e_x}$-axis.
   The latter is assumed to be aligned with the line of sight
   of the observation
   and the $\vec{e_z}$-axis with
   the cluster's outward radial direction
   (see Fig. \ref{fig:schema} for an illustrative sketch).
   Each direction is sampled with $128$ different cells,
   which are therefore not cubic
   but rather rectangular.
   Their volume is $\ell_x\times\ell_y\times\ell_z$,
   with $\ell_y{=}\ell_z{<}\ell_x$.

   Each cell is assumed to radiate in X-ray
   according to the APEC emission spectrum for collisionally ionised diffuse gas
   \citep{smith},
   with no galactic absorption considered:
   we checked that taking it into account with the PHABS model and a hydrogen column density
   of $5\times10^{20} \ \mathrm{cm}^{-2}$ leads to a decrease in flux only by a factor of two
   in the $0.2-1$ keV range and to almost no flux loss beyond $1$ keV.
   A synthetic observation is performed
   thanks to the E2E simulator SIXTE \citep{dauser},
   with a total exposure time of $\texp$ Ms.
   We refer to this procedure as the forward problem.

   Such a virtual observation can be reverse-engineered 
   by fitting the mock spectra in the $0.2-12$ keV range
   with the APEC model using the
   X-ray fitting package XSPEC,
   after gathering pixels into bins,
   in order to reconstruct the observed thermodynamic
   and velocity fields.
   This procedure is referred to as the inverse problem.

\begin{figure}
   \centering
   \includegraphics[width=\hsize]{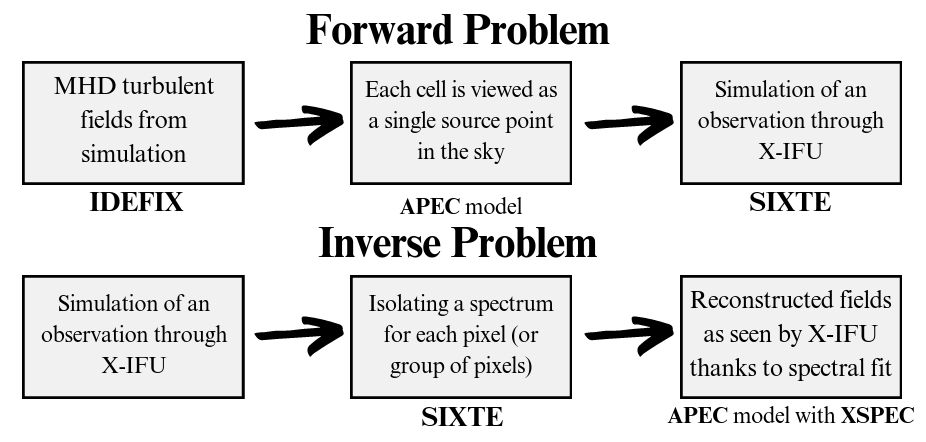}
      \caption{
              Schematic diagram of the pipeline to build a X-IFU synthetic
              observation and to reconstruct the physical fields from such an observation:
              the forward and inverse problems are solved separately.
              The tool and/or model used at each step is indicated
              below the respective boxes.
              }
      \label{fig:diagram}
   \end{figure}

   The whole pipeline (sketched in Fig. \ref{fig:diagram})
   is somewhat idealised because no background noise,
   no foreground absorption and no vignetting effects are included.
   It is however sufficient for the purpose of this exploratory study.
   More details about the technical work required to
   solve the forward and inverse problems can be found respectively
   in Appendix \ref{sec:forward} and \ref{sec:inverse}.

\section{Observational phenomenology of MTI turbulence in the ICM}
\label{sec:char}

   This section aims at obtaining a first phenomenological estimate 
   of the measurable properties of MTI-driven turbulence in the ICM
   as a function of the radial coordinate,
   in terms of both injection length and observable 2D turbulent
   kinetic and buoyancy potential energies.
   For this purpose, we make use of the three thermodynamic models
   presented in Section \ref{sec:prof}.
   Physical quantities in this section are dimensional
   and either deduced from the local values
   or from the local scale-heights
   of the aforementioned thermodynamics models at a given radius.
   Although those models are global in essence,
   here they are only used for their local properties (which however vary with radius).

   We first introduce the physics of the MTI and of its saturation mechanism.
   We then present scaling laws for the injection length and
   for the root-mean square of the 3D velocity and density fluctuation fields
   of MTI-induced turbulence, theoretically derived by \citetalias{perrone2D}.
   We finally assess, from an observational point of view,
   how much the intensity of the measurable 2D fluctuations
   is affected by the statistical
   cancellation of 3D fluctuations with opposite signs present along the line of sight.
   These first results will provide guidance on where to point the X-IFU
   if we were to detect features of magnetised turbulence induced by the MTI
   in ICM haloes.
   \subsection{Physics of the MTI}
   \label{sec:physics}
   The simplest MTI-favourable configuration consists in
   a dynamically negligible horizontal magnetic field threading
   a plasma layer with temperature gradient 
   and gravity both pointing downwards.
   When a blob of fluid is subsonically displaced
   from its initial equilibrium position,
   it is in pressure equilibrium with its new surrounding
   but remains thermally connected to its initial neighbours
   thanks to the dragging of the magnetic field lines.
   Provided that conduction 
   parallel to the magnetic field
   is fast enough,
   the perturbation is then isothermal.
   A rising blob will then
   be hotter (and less dense) than the neighbouring fluid and keep rising
   through buoyancy, triggering the MTI.

   There are three ways for this instability to be inhibited, namely through
   viscous damping, magnetic tension and entropy stratification in a stable atmosphere
   (with respect to the Schwarzschild criterion),
   which are specific of short (for the two first effects) and long wavelengths respectively.
   On scales in between, the MTI can fully develop up to a state of saturated turbulence
   in the ICM on timescales ${\sim}$Gyr, shorter than the Hubble time,
   with a maximal growth rate $\omega_T = \sqrt{g_0/H_T}$.
   The MTI way of saturating remains however unclear.
   \citetalias{perrone2D}
   argue that the instability saturates in such a way that
   the energy injected
   is roughly balanced by anisotropic thermal diffusion
   at all scales,
   leading to theoretical scaling laws that they were able to numerically
   verify for the kinetic and buoyancy potential energies as well as
   for the injection length.
   But other authors \citep{parrish2012a, mccourt}
   claim that MTI saturation is best described by a
   standard convection-like mixing-length theory, which would lead to 
   other dependencies of the kinetic energy on the local or global atmospheric model.
   It could be that the former saturation mechanism is best suited to describe
   MTI turbulence on scales smaller than typical ICM scale-heights,
   while the latter mixing-length theory would describe MTI turbulence
   on larger scales, with a transition regime occurring in regions where the
   instability length scale becomes of the order of the temperature scale-height.  
   In this paper, we make intensive use of the scaling laws from \citetalias{perrone2D},
   therefore fostering a local approach to MTI turbulence,
   which we subsequently find to be justified for ICM radii sufficiently small
   where X-ray emissivity is high enough
   to make observations possible (see Section \ref{sec:length});
   however, global models are needed to eventually settle this debate.

   \subsection{Length scale of the MTI}
   \label{sec:length}
\begin{figure}
   \centering
   \includegraphics[width=\hsize]{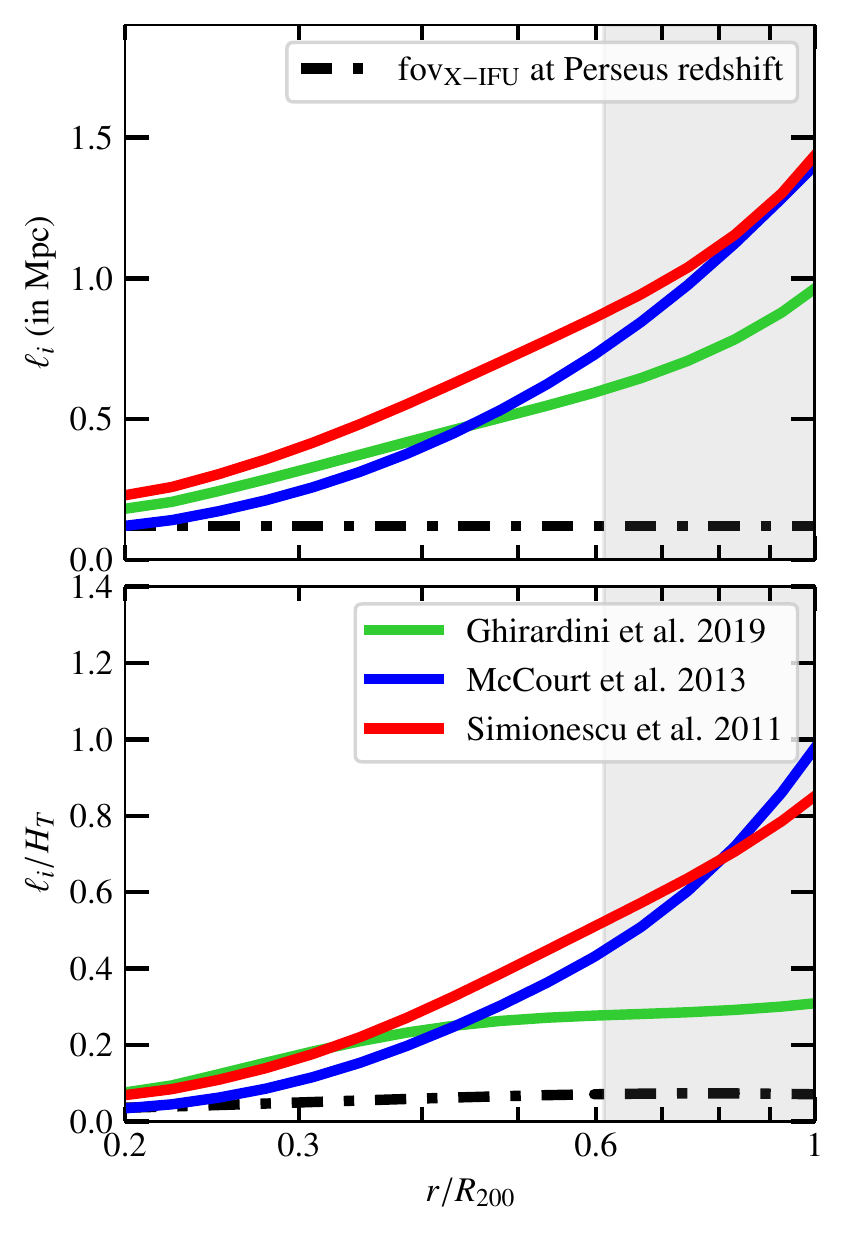}
      \caption{Theoretical injection length of turbulence induced by the MTI (top)
               and the same quantity divided
               by the local temperature scale-height (bottom)
               as a function of the radius
               for three models of ICM.
               The colour encoding is
               the same as in Fig. \ref{fig:profiles}.
               The dash-dotted black line is the X-IFU
               field of view at Perseus redshift.
               Again, the grey shaded areas highlight radii
               at which a X-IFU observation will not provide enough photons.
               These radii also roughly correspond to regions
               where the hypothesis of locality for
               MTI-driven turbulence may break down
               (and so would the scalings Eqs. (\ref{eq:li})-(\ref{eq:K})-(\ref{eq:U})).
              }
         \label{fig:lengths}
   \end{figure}

   The scaling relation derived by \citetalias{perrone2D} for the injection length
   of MTI-driven turbulence reads:
\begin{equation}
    \ell_i \approx 1.1 \frac{(\chi \omega_T)^\frac{1}{2}}{N},
\label{eq:li}
\end{equation}
   with $N$ the Brunt-Väisälä frequency of a perfect gas
   and $\chi$ the Spitzer thermal
   diffusivity\footnote{In \citetalias{perrone2D}, the thermal diffusivity
   absorbs an additional $\left(\gamma-1\right)/\gamma$ factor
   with respect to the classical expression
   Eq. (\ref{eq:spitzer}).
   In the current work, this physical prefactor is included
   in the numerical prefactor of the scaling
   laws Eqs. (\ref{eq:li})-(\ref{eq:K})-(\ref{eq:U}).
   No suppression factor is considered.
   } \citep{spitzer}
   respectively defined according to:
\begin{equation}
   N = \sqrt{\frac{g_0}{\gamma H_S}},
\label{eq:N}
\end{equation}
\begin{equation}
   \chi = 4.98\times10^{31} \left( \frac{k_\mathrm{B} T}{5 \ \mathrm{keV}} \right)^{\frac{5}{2}} \left(\frac{n_\mathrm{e}}{10^{-3} \ \mathrm{cm^{-3}}} \right)^{-1} \ \mathrm{cm^2 s^{-1}},
\label{eq:spitzer}
\end{equation}
   where $n_\mathrm{e}$ is the electron number density.
   In \citetalias{perrone2D}, Eq. (\ref{eq:li}) is given without the numerical
   prefactor but the tabulated data are sufficient to recover it
   (its value was confirmed by the corresponding author, priv. comm.).
   The MTI injection length is not always small compared to 
   $H_T$,
   especially beyond $0.6R_{200}$
   and this may prevent us from using local models
   to study MTI dynamics in these regions.
   We consider that the use of local simulations remains justified to model 
   MTI dynamics at radii below
   $0.6R_{200}$, as
   $\ell_{i} {\lesssim} 0.5H_T$
   there for the three ICM models.
   The regions where this criterion is no longer fulfilled
   are located at large cluster radii,
   where the previously introduced criterion on the minimal number of photons
   required for a given exposure time is not met anyway
   (grey shaded areas in Figs. \ref{fig:profiles}-\ref{fig:lengths}-\ref{fig:3Drms}).
   This further justifies our local approach to MTI turbulence,
   and the use of the scaling laws from \citetalias{perrone2D}.
   We note that the expected injection length of MTI-driven turbulence
   at $0.6R_{200}$ is ${\sim}0.7$ Mpc, which
   is not negligible compared to the injection length typically associated with turbulence
   driven by mergers, accretion, shocks, etc.
   In the next subsection, we assess the
   levels of turbulence induced by the instability.

   \subsection{Turbulence levels of the MTI}
   \label{sec:level}
   To assess the expected levels of turbulence
   driven by the MTI in the ICM, we use the scaling relations
   for the root-mean square velocity
   $\left.\varv\right|_\mathrm{rms}$
   and density fluctuations
   $\left. \delta \rho \right|_\mathrm{rms}$
   derived in
   \citetalias{perrone2D}:

\begin{equation}
    \left. \varv\right|_\mathrm{rms}^2 \approx 0.024 \frac{\chi\omega_T^3}{N^2},
\label{eq:K}
\end{equation}
\begin{equation}
    \frac{g_0^2}{2N^2} \left. \delta \rho \right|^2_\mathrm{rms}  \left(\frac{N}{\omega_T}\right)^{\frac32} \propto \left(\frac{\chi\omega_T^3}{N^2}\right)^\frac{3}{4}.
\label{eq:Upropto}
\end{equation}

   While the former relation is provided as such in their paper
   and dimensionally consistent,
   the latter is provided without the prefactor, which must be
   dimensional in this case because of the weaker
   (to the power $3/4$)
   dependency of the buoyancy potential energy on $\chi\omega_T^3/N^2$
   making the right-hand side of Eq. (\ref{eq:Upropto}) not the dimension
   of a specific energy.
   This means that the turbulent saturation of this quantity is controlled
   by one or several other physical processes as resistivity or viscosity
   for instance (Perrone, priv. comm.).
   Such residual dependencies on both the magnetic and the usual Prandtl numbers
   ($\mathrm{Pm}=\nu/\eta$ and $\mathrm{Pr}=\nu/\chi$ respectively)
   have been numerically highlighted at least in 2D
\defcitealias{perrone2D}{PL22a}
   \citepalias{perrone2D}.
\defcitealias{perrone2D}{PL22a,b}
   Further theoretical and numerical work
   would be needed to overcome this limitation
   and to derive the full dependencies.
   Absent such a fully self-consistent scaling law,
   we chose to dimensionalise the numerical prefactor,
   which we determined to be $0.003$ thanks to the data available in \citetalias{perrone3D},
   by the root-square of the thermal velocity $\varv_\mathrm{th}$
   at $k_\mathrm{B}T{=}5$ keV.
   This is the easiest way of making
   this scaling dimensionally consistent without using
   any
   MTI-related quantities,
   which would otherwise introduce unwelcome dependencies
   of the potential energy
   on MTI-related parameters
   that are not seen in the results from \citetalias{perrone2D}.
   The scaling relation
   for the buoyancy potential energy finally reads:

\begin{equation}
   \frac{g_0^2}{2N^2} \left. \delta \rho \right|^2_\mathrm{rms} \left(\frac{N}{\omega_T}\right)^{\frac32} \approx 28\ \mathrm{\left(cm/s\right)}^{\frac12} \times \left(\frac{\chi\omega_T^3}{N^2}\right)^\frac{3}{4}.
\label{eq:U}
\end{equation}
   \begin{figure}
   \centering
   \includegraphics[width=0.9\hsize]{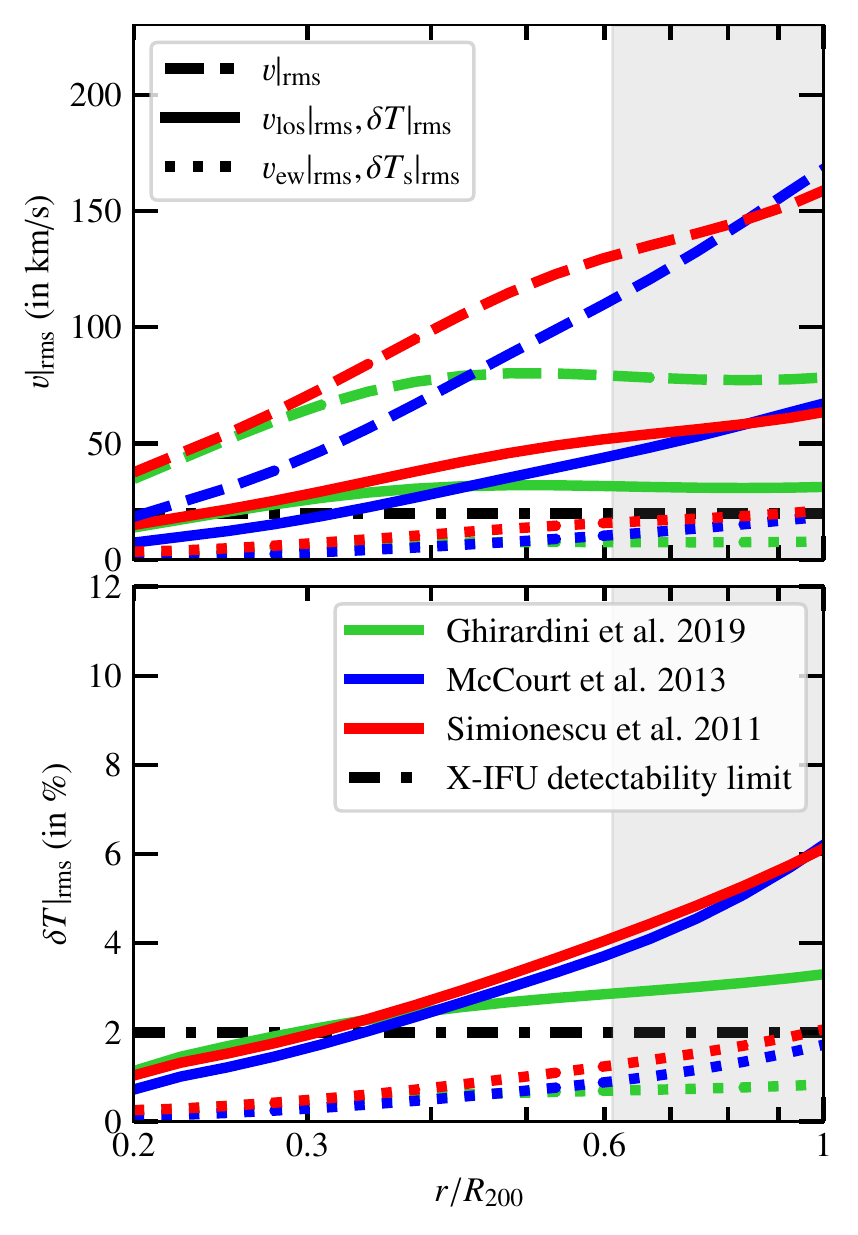}
      \caption{Intensity levels of MTI turbulence.
               Dashed line: root-mean square of the 3D turbulent
               velocity field as a function of the radius in the ICM.
               Full line: root-mean square of the 3D turbulent
               line-of-sight velocity (top)
               and temperature (bottom) fluctuations.
               Dotted line: same but for the estimated 2D projected quantities
               after integration along the line of sight.
               The dash-dotted black lines are X-IFU
               detectability limits
               (determined in Section \ref{sec:results})
               for a typical $\texp$-Ms
               observation of a Perseus-like cluster at 
               $\x R_{200}$.
               The colour encoding follows Fig. \ref{fig:profiles}.}
         \label{fig:3Drms}
   \end{figure}

   Eqs. (\ref{eq:K})-(\ref{eq:U}) translate into moderate values
   of the root-mean square velocity
   $\mathcal{O}(100\ \mathrm{km/s})$
   (dashed lines on the top plot in Fig. \ref{fig:3Drms})
   and temperature fluctuations\footnote{MTI-induced turbulence is expected to be subsonic.
             If so,
             the density and temperature fluctuations
             are anti-correlated and the same
             scaling law can be used for both fluctuation fields,
             as $\left. \delta \rho \right|_\mathrm{rms} = \left. \delta T \right|_\mathrm{rms}$.
             }
   $\mathcal{O}(5 \%)$
   (full lines on the bottom plot),
   in the observable regions of interest.
   According to \citetalias{perrone2D},
   the MTI turbulent velocity field at saturation
   exhibits an anisotropy with respect to the direction
   of gravity;
   with the vertical (i.e. radial) component of the velocity being larger
   than its horizontal (i.e. azimuthal) counterparts.
   For illustrative purposes,
   we consider here that
   the root-mean square of the line-of-sight velocity field
   (full lines on the top plot in Fig. \ref{fig:3Drms})
   is smaller than the root-mean square velocity itself by a factor ${\sim}\ani$
   (this value will be further justified by the simulation in Section \ref{sec:results}),
   rather than the usual $\sqrt{3}$ factor for isotropic turbulence.
   Indeed, X-ray spectrometers like the X-IFU have only access
   to the line-of-sight velocity
   because it is deduced from the measurement of the Doppler shift of spectral lines.
   The relative contribution of the horizontal component of the velocity field
   to the line-of-sight velocity
   increases as the X-IFU points further away from the cluster centre, as illustrated in Fig. \ref{fig:schema}.
   Indeed, at large ICM radii, the projection of the vertical (i.e. radial) velocity
   on the line-of-sight direction is weak,
   especially at locations where the plasma emissivity is high.
   This geometrical effect has two unfavourable consequences
   for the possible detection of the MTI.
   On the one hand, the anisotropy of the kinetic energy
   components makes the observed line-of-sight velocity,
   which mostly correlates with the horizontal velocity in cluster haloes,
   weaker than its purely vertical counterpart and therefore harder to measure.
   On the other hand, the horizontal velocity
   does not correlate with density fluctuations;
   unlike the vertical velocity,
   due to the convective nature of the instability.
   Combined together, these two effects might limit the opportunity to
   detect relevant signatures of the MTI in the ICM.
   A naive solution to observe the vertical component of the velocity field would be to
   look at the ICM arbitrarily close to its centre.
   However, this is not possible
   since we are aiming at regions with decreasing background temperature
   in relaxed CC clusters,
   and central regions are likely more disturbed by AGNs.

   \subsection{Observation along the line of sight}
   \label{sec:projection}
   The values of the root-mean square fluctuations
   shown in Fig. \ref{fig:3Drms} (full lines) are for the 3D line-of-sight velocity
   and the 3D temperature fluctuation fields.
   However, these 3D fields are not available as such from an
   observation but their 2D counterparts 
   integrated along the line of sight are because the ICM is optically thin.
   More precisely, the observed line-of-sight velocity is often assimilated 
   to the emission-weighted velocity
   \citep{roncarelli,cucchetti}
   and the observed temperature
   to the spectroscopic temperature
   \citep{mazzotta},
   respectively defined as:
\begin{equation}
   \varv_\mathrm{ew} = \left. \int \rho^2 \varv_\mathrm{los} ds \right/\int \rho^2 ds,
   \label{eq:vew}
\end{equation}
\begin{equation}
   T_\mathrm{s} = \left. \int \rho^2 T^{0.25} ds\right/\int \rho^2 T^{-0.75} ds,
   \label{eq:Ts}
\end{equation}
   where $s$ is the line-of-sight coordinate (Fig. \ref{fig:schema}).
   Accordingly and for later use,
   we also define the emission measure which tracks the total plasma emissivity:
\begin{equation}
   EM = \int \rho^2 ds.
   \label{eq:EM}
\end{equation}
   When observing or integrating along the line of sight, fluctuations with opposite signs will
   statistically cancel each other, especially when the
   average size of turbulent eddies is small
   with respect to the length over which significant emission takes place
   (which is precisely the length over which the fields should be integrated).
   We now aim at quantifying the reduction in signal due to this observational effect,
   illustrated in Fig. \ref{fig:schema}.
   
   Hints about the behaviour of the
   root-mean square and dispersion of the emission-weighted velocity
   and of the second-order velocity structure function
   after integration along the line of sight
   can be gained from previous works by
   \citet{churazov2012}, \citet{zhuravleva2012}, \citet{zuhone}, \citet{clerc}, and \citet{mohapatra22}.
   Many different models have been developed,
   accounting for the two-dimensionality of the observable fields
   or the variation of the background thermodynamic fields over the field of view for example.
   The present phenomenological discussion
   does not require such sophisticated derivations. So we chose the simplest model,
   namely the root-mean square velocity along a single line of sight
   (Eq. (5) from \citealt{clerc},
   which is identical to Eq. (A9) in \citealt{zhuravleva2012}),
   to infer the decrease in signal due to fluctuation-cancellation along the line of sight.
   In both equations, the brackets denote an average over independent realisations
   of a 1D velocity field with the same underlying power spectra.
   In our case, the assumption made when using this simple model is therefore to assume that
   the root-mean square of a single realisation of an emission-weighted
   2D velocity field
   is equivalent to
   the root-mean square of independent realisations of a 1D velocity field, 
   postulating the same underlying power spectra for all processes.
   Although this hypothesis may not hold in general,
   we deem the estimates obtained from it
   sufficiently enlightening as a zeroth-order assessment of the signal reduction
   due to the cancellation of fluctuations along the line of sight.
   Using this toy-model for the different density profiles
   (with the MTI turbulent energy spectra that will be later presented
   in Section \ref{sec:results}, Fig. \ref{fig:spec}),
   we obtain
   first estimates of the root-mean square
   emissivity-weighted velocity after
   integration along the line of sight for MTI-induced turbulence.
   In the case of temperature fluctuations, the spectroscopic weighting scheme introduces
   an additional complexity with respect to the simple emissivity weighting.
   However we checked, thanks to simulated turbulent fields,
   that the levels of fluctuation-cancellation of temperature fluctuations
   are similar for both weighting schemes.
   We therefore use the same toy-model
   Eq. (5) from \citealt{clerc}
   to anticipate the loss of signal due to the cancellation of fluctuations
   along the line of sight
   for both the emissivity-weighted velocity field
   and the spectroscopic temperature fluctuations.
   These forecasts are shown as dotted lines in Fig. \ref{fig:3Drms}.
   
   The results of this section are well summarised in Fig. \ref{fig:3Drms}.
   The latter overall shows that the root-mean square
   of the observable spectroscopic temperature fluctuations
   (dotted lines on the bottom panel) is below 
   the root-mean square of the respective 3D temperature fluctuation field
   (full lines on the same plot).
   The only effect responsible for that 
   is the cancellation of opposite-sign fluctuations
   when observing along the line of sight.
   The situation is slightly less straightforward in the case of the velocity field.
   The MTI-driven turbulent 3D velocity field has a pristine root-mean square
   intensity deduced from Eq. (\ref{eq:K})
   (dashed lines on the top panel in Fig. \ref{fig:3Drms}).
   However, only the line-of-sight component of the velocity field is accessible
   to an observation. In cluster peripheries, the line of sight is mostly aligned with the
   azimuthal direction rather than the radial direction
   (at least wherever the plasma emissivity is higher).
   As shown in \citetalias{perrone2D},
   the MTI saturates
   in such a way that the vertical component of its velocity field
   carries most of the total kinetic energy.
   As a result, the root-mean square of the 3D line-of-sight velocity field
   (before integration along the line of sight, full lines on the top plot)
   is ${\sim}2.5$ smaller
   than the pristine root-mean square of the full 3D velocity field
   (dashed lines on the top plot in Fig. \ref{fig:3Drms}).
   Then, the cancellation of fluctuations along the line of sight brings further
   down the root-mean square of the emissivity-weighted velocity field
   (dotted lines on the top panel in Fig. \ref{fig:3Drms}).
   Our analysis suggests that the estimated ratio
   between the actual root-mean square of a 3D MTI-like fluctuation field
   and its measurable line-of-sight integrated 2D counterpart ranges
   from 2.6 to 5.6
   according to the density profile used and the ICM radial coordinate.
   Keeping this phenomenology in mind, we assess
   in Section \ref{sec:results}
   whether such moderate levels
   of fluctuations
   ($\left.\varv_\mathrm{ew}\right|_\mathrm{rms}{\sim}10\ \mathrm{km/s}, \ \left.\delta T_\mathrm{s}\right|_\mathrm{rms}{\sim}0.5\%$
   at $\x R_{200}$ in a Perseus-like cluster)
   would be detectable with our baseline instrumental X-IFU configuration
   (the corresponding detectability limits are shown in Fig. \ref{fig:3Drms}),
   by modelling the observable ICM magnetised dynamics more quantitatively
   thanks to a Braginskii-MHD simulation and to X-IFU synthetic observations.

\section{Quantitative numerical study with virtual observations}
\label{sec:results}
   Armed with the previous phenomenological estimates,
   we now characterise more precisely
   the statistical properties and the spatial structure of
   MTI dynamics in the ICM
   and assess to what extent these characteristics may be recovered
   from an observation through the X-IFU.
   For this purpose, we first present
   diagnostics derived from a local Braginskii-MHD numerical simulation of MTI turbulence.
   We then discuss the reason why the saturation levels seen in the simulation
   are not representative of those found in the real ICM
   by comparing the numerical against ICM plasma regimes,
   and we rescale the turbulent fluctuation fields of the simulation accordingly.
   Finally, we perform two X-IFU mock observations of a snapshot of the rescaled MHD simulation
   in the stage of developed turbulence.
   We evaluate
   the quality of the reconstructed fields for each run
   and we bring to light
   the inherent anisotropy of MTI-induced turbulence
   in the second mock observation.

   \begin{figure}
   \centering
   \includegraphics[width=0.93\hsize]{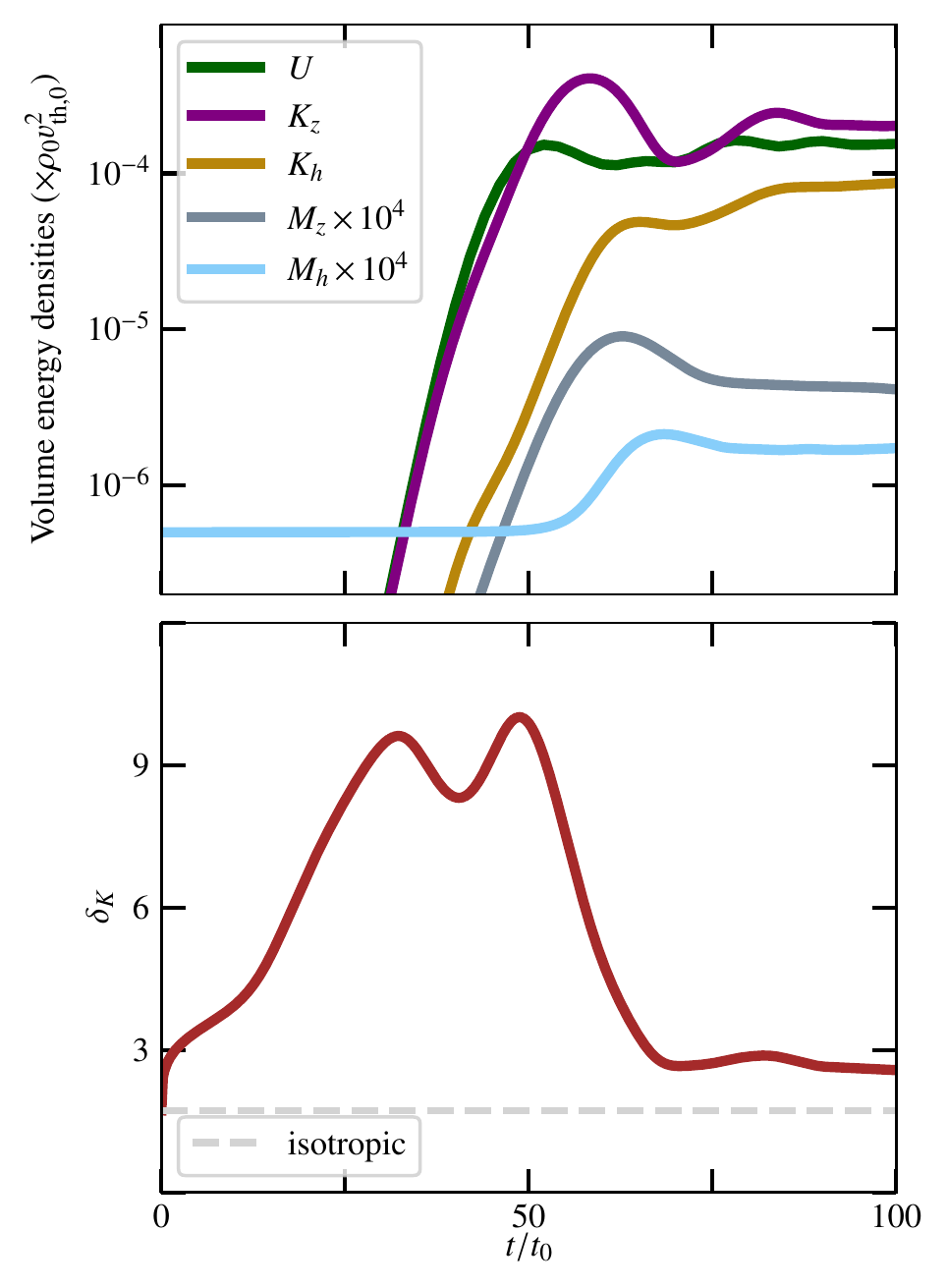}
      \caption{Time evolution of key MTI quantities.
               Top: Potential (in green), vertical (in dark purple), and
               horizontal (in gold) kinetic
               energy densities and vertical (in dark blue)
               and horizontal (in light blue) magnetic energy densities.
               Bottom: Kinetic energy anisotropy
               $\delta_K = \sqrt{2K/K_h}$.
               }
      \label{fig:time}
   \end{figure}

   \begin{figure*}
   \centering
   \includegraphics[width=\hsize]{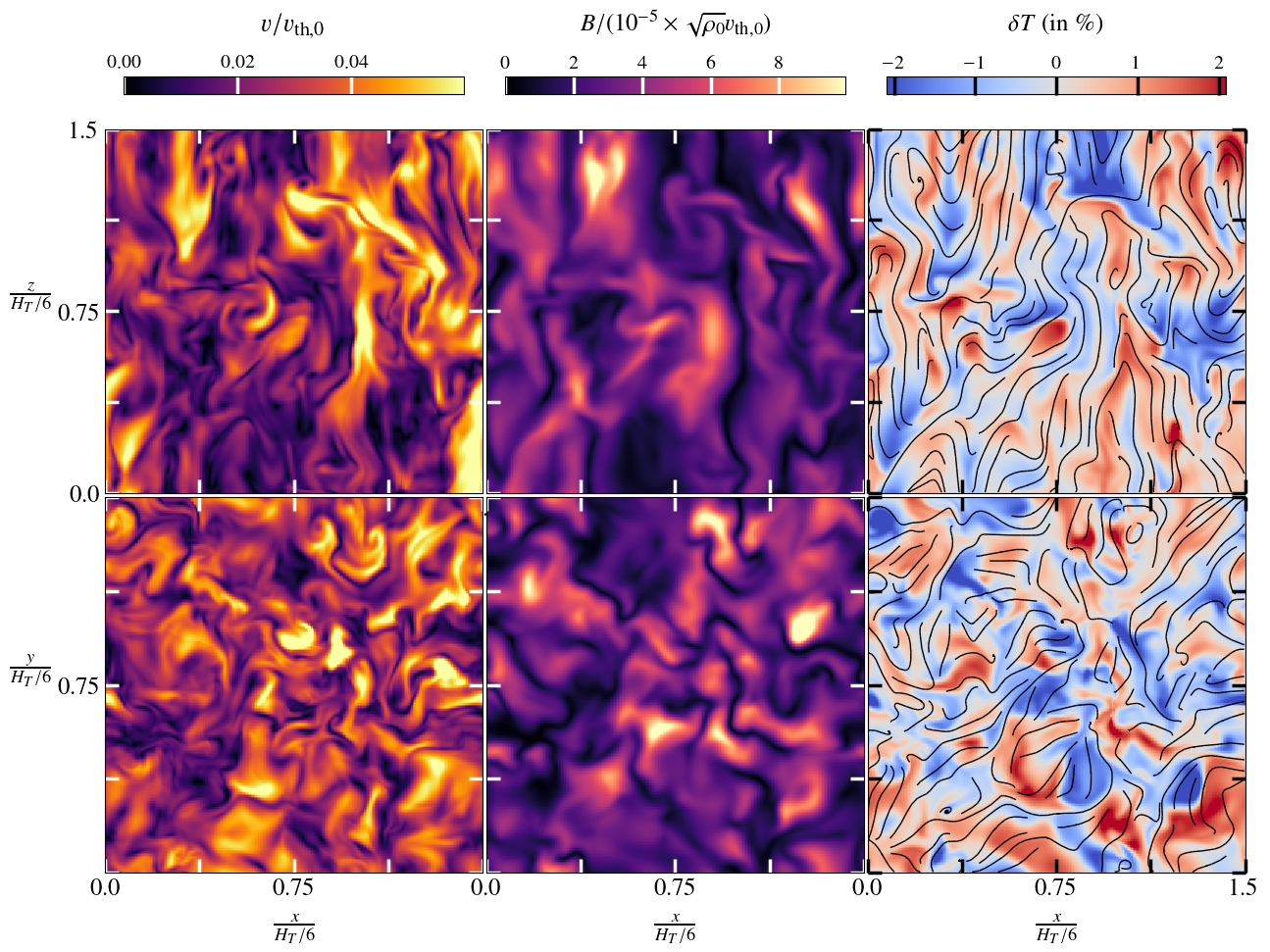}
      \caption{Snapshots of MTI turbulent fields
               at $t=100t_0$.
               Top: Vertical $(x,z)$ cuts at $y=0.75$.
               Bottom: Horizontal $(x,y)$ cuts at $z=0.75$.
               From left to right:
               Norm of the velocity and magnetic fields 
               (in code units)
               and temperature fluctuations
               (in $\%$ of the background temperature $T(z)$, Eq. (\ref{eq:Tz})).
               Magnetic field lines
               are superimposed as black lines on the temperature fluctuations.
               }
      \label{fig:slices}
   \end{figure*}

\subsection{Simulation of MTI-driven turbulence}
   The time evolution of key quantities related to the MTI
   are shown in Fig. \ref{fig:time}.
   In the top panel, the average volume density
   of the potential, kinetic and magnetic energies
   during the simulation, respectively defined as
   $U = g_0^2/(2N^2){\times}{<\rho\delta\rho^2}>$,
   $K = {<\rho \varv^2>}/2$ and $M = {<B^2>}/2$,
   are presented.
   In the linear regime,
   the MTI displays an exponential growth phase
   before saturating in a turbulent state.
   The anisotropy between the vertical
   and horizontal
   components of the kinetic energy
   is quantified by the bias
   $\delta_K = \sqrt{2K/K_h}$, with $K_h=K_x+K_y$ (bottom panel in Fig. \ref{fig:time}),
   which is higher during the linear growth phase than at saturation,
   where it settles around \ani.
   This is the value we used to convert from $\left.\varv\right|_\mathrm{rms}$ to $\left.\varv_\mathrm{los}\right|_\mathrm{rms}$ in Fig. \ref{fig:3Drms}
   introducing the qualitative phenomenology of MTI observations.
   This strong anisotropy
   is a well-known feature of the MTI \citepalias{perrone2D}.
   The instability is indeed driven by the buoyancy
   and tends to fight its physical origin, namely
   the temperature gradient.
   The latter is brought to isothermality via
   preferentially vertical convection-like fluid motions
   though the anisotropic conductive heat flux also comes into play in this process.
   We note that
   Braginskii-MHD equations with anisotropic viscosity support
   a fluid version of the firehose instability \citep{kunz}.
   We do not see such a small-scale instability develop in our simulation,
   certainly on account of the numerical dissipation at the grid level
   and of the low magnetic Reynolds number of the simulation.

   Although the magnetic field initially grows as a result of the
   MTI, the magnetic field evolution
   shows no sign of an exponential kinematic dynamo phase after the MTI saturates
   because of the value of the magnetic Reynolds number,
   which is about $5.4$ in our simulation.
   This is below the critical magnetic Reynolds number
   $\mathrm{Rm_c}=35$
   identified by \citetalias{perrone3D}
   for the small-scale fluctuation dynamo to initiate
   a kinematic growth with MTI turbulence.
   Therefore, the magnetic field strength here never reaches equipartition
   with the turbulent kinetic energy;
   unlike in the higher-resolution fiducial run from \citetalias{perrone3D} which has $\mathrm{Rm}=110$.

   Fig. \ref{fig:slices} shows key MTI turbulent fields
   at the very end of the simulation (at $t{=}\tend t_0$):
   vertical and horizontal snapshots of the velocity and magnetic field norms,
   and of the fluctuation temperature field.
   Because the magnetic and velocity fields are divergence-free,
   the vertical anisotropy related to the different components of the magnetic
   and kinetic energies reflects into the shape of the turbulent structures which are
   more elongated along the vertical direction than in the horizontal directions
   (leftmost and centre slices in Fig. \ref{fig:slices}).
   Such an anisotropy is also seen in the temperature fluctuation field (rightmost slices).
   Another interesting feature of anisotropic heat
   diffusion is the formation of
   almost isothermal magnetic field lines
   with possibly strong temperature gradients across them
   (rightmost slices in Fig. \ref{fig:slices}).
   Observing such temperature fluctuations in the
   ICM could thus indirectly shed light on the magnetic field geometry,
   if anisotropic heat flux is active in such plasma.

\begin{figure}
   \centering
   \includegraphics[width=\hsize]{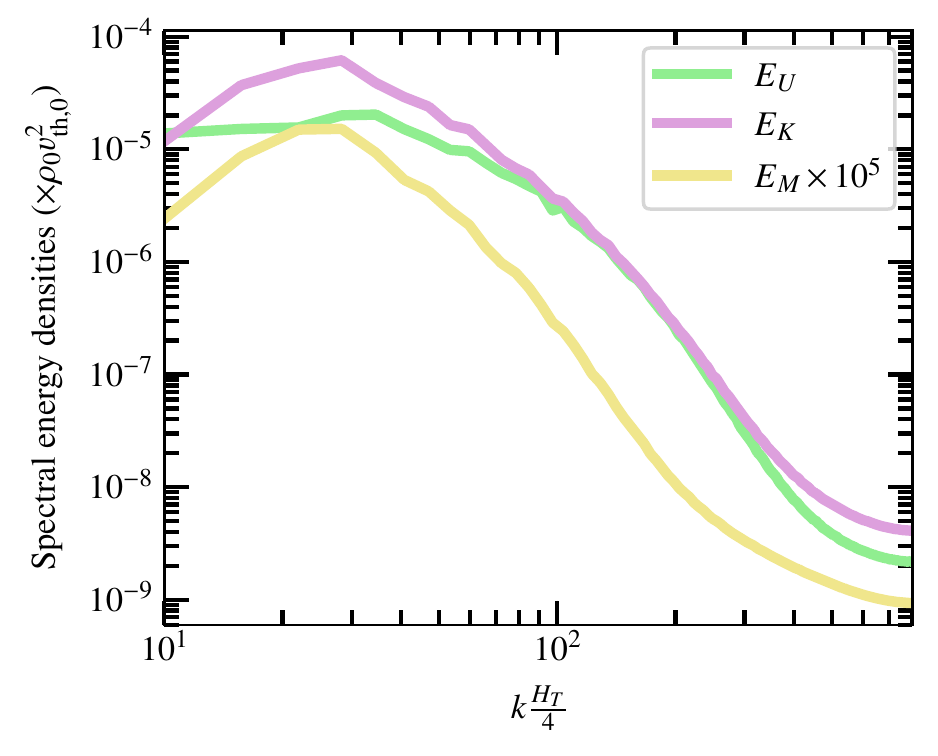}
      \caption{Energy spectra of
              the potential (green),
              kinetic (purple)
              and magnetic (yellow)
              energy volume densities.
              }
      \label{fig:spec}
   \end{figure}

   In Fig. \ref{fig:spec}, we look at the shell-integrated spectral energy volume densities
   at saturation (at $t{=}\tend t_0$).
   In the case of the kinetic energy, this quantity is:
\begin{equation}
   E_K (k) = \sum_{k\le|\vec{k}|\le k + \Delta k} \frac12 \sum_{i=x,y,z} \left| \mathcal{F}[\sqrt{\rho} \varv_i](\vec{k})\right|^2,
\end{equation}
   where $\sqrt{\rho}\varv_i$ needs to be replaced by $B_i$
   for computing the spectra of the magnetic energy.
   For the potential energy, it is:
\begin{equation}
   E_U (k) = \sum_{k\le|\vec{k}|\le k + \Delta k} \frac{g_0^2}{2N^2} \left| \mathcal{F}\left[\sqrt{\rho}\delta \rho\right](\vec{k})\right|^2.
\end{equation}
   $\mathcal{F}$ is the Fourier transform of the field between brackets
   and $\vec{k}=k_x\vec{e_x}+k_y\vec{e_y}+k_z\vec{e_z}$ the wavevector.
   Taking the Fourier transform of the previous fields
   is justified here for the $x$ and $y$ coordinates
   because the BC are periodic in these directions.
   This is also justified in the vertical $z$ direction here thanks to the quasi-periodic BC
   which periodises the velocity and magnetic field and,
   in the case of the thermodynamic fields,
   their fluctuations rather than the fields themselves.
   All fields present a typical phenomenological picture of developed
   turbulence in which energy cascades from
   larger to smaller
   scales
   in spectral space,
   though on a relatively small range of scales here.
   The large-scale flow is characterised by the 
   injection scale
   $\ell_i$,
   defined as
\begin{equation}
   \ell_i = \frac{2\pi}{k_i} = 2\pi \frac{\int k^{-1} E_K dk}{\int E_K dk}.
\end{equation}
   In the MHD simulation, we have $k_i \approx \ksisim L^{-1}$, meaning that the injection length
   is about $\elliL \%$ of the box length.
   The turnover time $\ell_i/\varv_\mathrm{rms}$ is therefore ${\sim}13t_0$.

\subsection{Rescaling of the simulation}
\label{sec:rescaling}
   We aim at making virtual observations at $\x R_{200}$
   in a Perseus-like cluster but
   the dimensional parallel heat diffusivity used in the simulation
   $\chi=\diff \varv_{\mathrm{th},0} H_T/6$
   is roughly two orders of magnitude less than the value expected
   from the Spitzer diffusivity
   Eq. (\ref{eq:spitzer})
   at the same radius.
   We therefore anticipate, from the scaling relation Eq. (\ref{eq:K}),
   that the raw dimensional root-mean square velocity in our MHD simulation
   is one order of magnitude
   less than what would be found in the ICM.
   The same argument holds for the potential energy.
   For instance, Eq. (\ref{eq:K}) is equal to
   $14$ km/s
   when used with the simulation parameters given in \ref{sec:config}
   and then dimensionalise with the thermal velocity $\varv_{\mathrm{th},0}$
   of the Perseus cluster at $\x R_{200}$.
   But Eq. (\ref{eq:K}) can also directly be used with the physical
   parameters deduced from Perseus thermodynamic profiles at the same radius
   as done in Section \ref{sec:char}, Fig. \ref{fig:3Drms},
   in which case we find $\left.\varv\right|_\mathrm{rms}{\sim}55$ km/s.
   Consequently, to make a virtual observation,
   the fluctuation fields of the simulation first need to be
   rescaled according to Eqs. (\ref{eq:K})-(\ref{eq:U}) to match
   the expected intensity of MTI-induced turbulent fluctuations
   in the realistic ICM diffusion regime.
   Similarly, the scaling law Eq. (\ref{eq:li}) predicts
   the physical injection length of MTI turbulence in the ICM,
   which is not the one that we see in the simulation
   (when the lengths are dimensionalised by $H_T/6$)
   since the simulation is not in the right regime.
   Eq. (\ref{eq:li}) thus provides a new length scale from which
   the box size of the MHD simulation can be dimensionalised,
   by matching the expected physical injection length with
   the one found in our simulation.

   This physically motivated rescaling is always conducted
   before any synthetic observation is performed in the next section.
   A peculiar rescaling is fully characterised by
   the final injection length $\ell_i$ and root-mean square
   of the emission-weighted velocity
   $\left. \varv_\mathrm{ew} \right|_\mathrm{rms}$
   and spectroscopic temperature fluctuations $\left. \delta T_\mathrm{s} \right|_\mathrm{rms}$.

   \begin{figure*}
   \centering
   \includegraphics[width=\hsize]{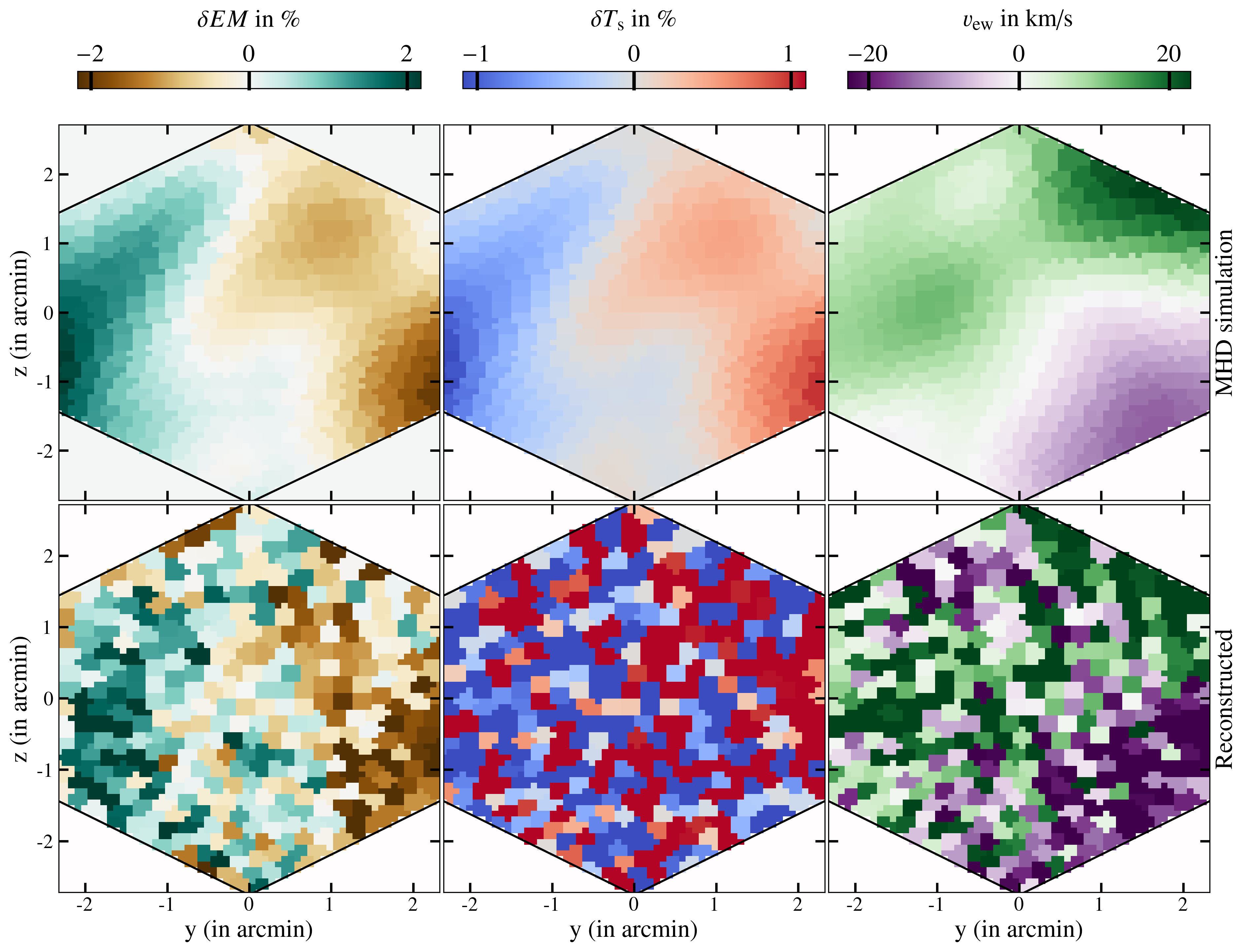}
      \caption{
              The fiducial synthetic observation OBSfid based on
              the Braginskii-MHD simulation rescaled with the values of
              Eqs. (\ref{eq:li})-(\ref{eq:K})-(\ref{eq:U})
              at $\x R_{200}$ in a Perseus-like cluster.
              From left to right: emission measure fluctuation $\delta EM$,
              spectroscopic temperature fluctuation $\delta T_\mathrm{s}$
              and emission-weighted velocity $\varv_\mathrm{ew}$ maps.
              Top: line-of-sight integrated input fields from the rescaled MHD simulation.
              Bottom: output fields reconstructed from the synthetic observation OBSfid.
              The colourbars are set according to the input quantities.
              The values of the thermodynamic fluctuation fields are given
              in percents of the average background quantities
              (see Eq. (\ref{eq:flucV})).
              }
      \label{fig:obsfid}
   \end{figure*}

   \begin{figure*}
   \centering
   \includegraphics[width=\hsize]{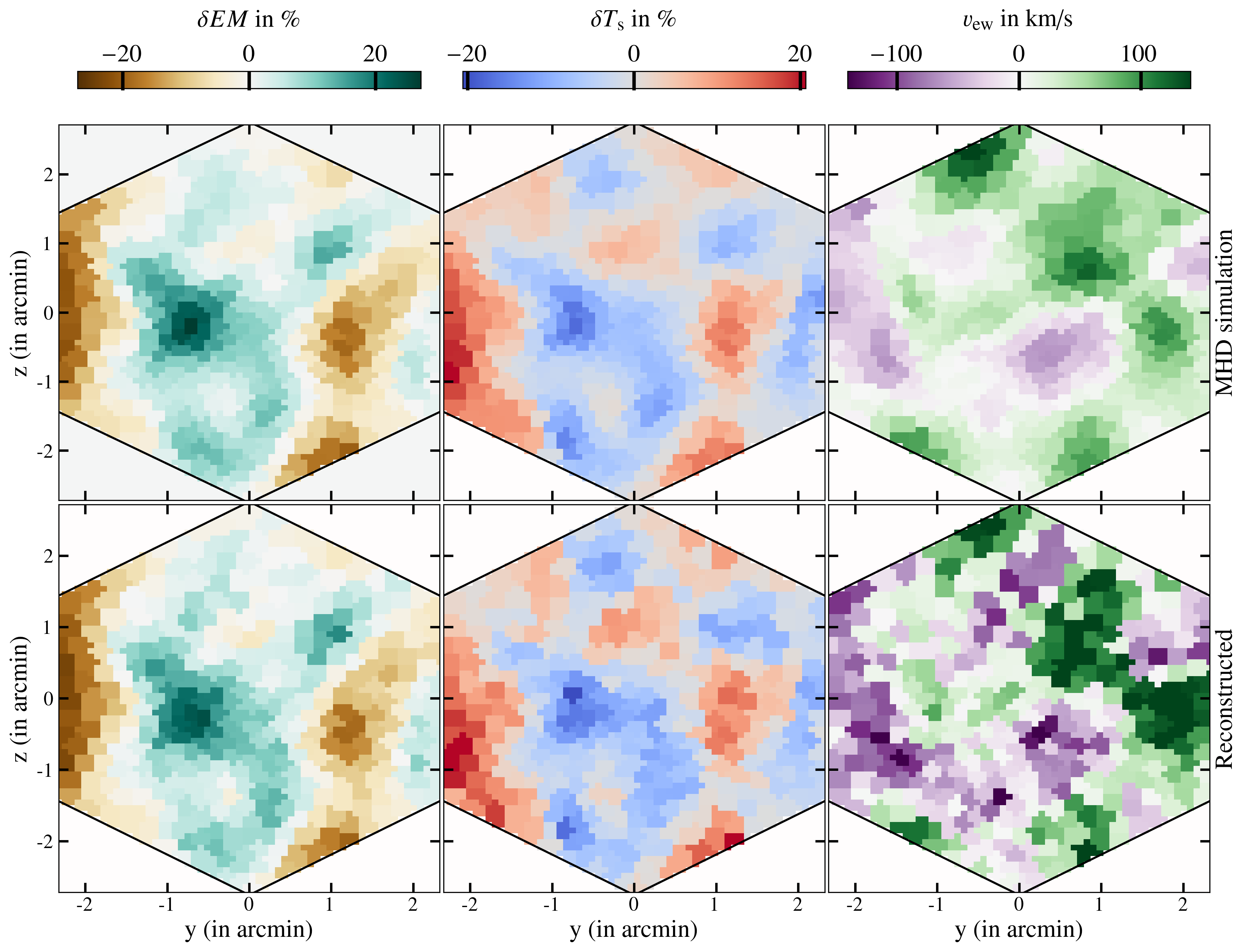}
      \caption{
              The custom synthetic observation OBScustom
              (at $\x R_{200}$ with Perseus density profile),
              based on the optimally custom-rescaled Braginskii-MHD simulation.
              From left to right: emission measure fluctuation $\delta EM$,
              spectroscopic temperature fluctuation $\delta T_\mathrm{s}$
              and emission-weighted velocity $\varv_\mathrm{ew}$ maps.
              Top: line-of-sight integrated input fields from the rescaled MHD simulation.
              Bottom: output fields reconstructed from the synthetic observation OBScustom.
              The colourbars are set according to the input quantities.
              The values of the thermodynamic fluctuation fields are
              in percents of the average background quantities
              (see Eq. (\ref{eq:flucV})).
              }
      \label{fig:obscustom}
   \end{figure*}

\subsection{Virtual observations of MTI-induced turbulence with the X-IFU}
   We now consider the final state (after ${\sim}8$ turnover times)
   of the MHD simulation
   examined in the previous sections,
   and virtually observe its rescaled dynamics
   through the X-IFU according to the methodology described in Section \ref{sec:synthetic}.
   Solving the forward problem, we obtain a mock observation,
   which we can then reverse-engineer by solving the inverse problem
   to reconstruct the emission measure, temperature and velocity fields.
   In the next sections,
   we display the maps derived from the 
   synthetic observations.
   We qualitatively (and quantitatively in the Appendix \ref{app:analysis}) compare them
   to the respective pristine input data from the rescaled Braginskii-MHD simulation,
   numerically integrated along the line of sight.

   \begin{figure*}
   \centering
   \includegraphics[width=\hsize]{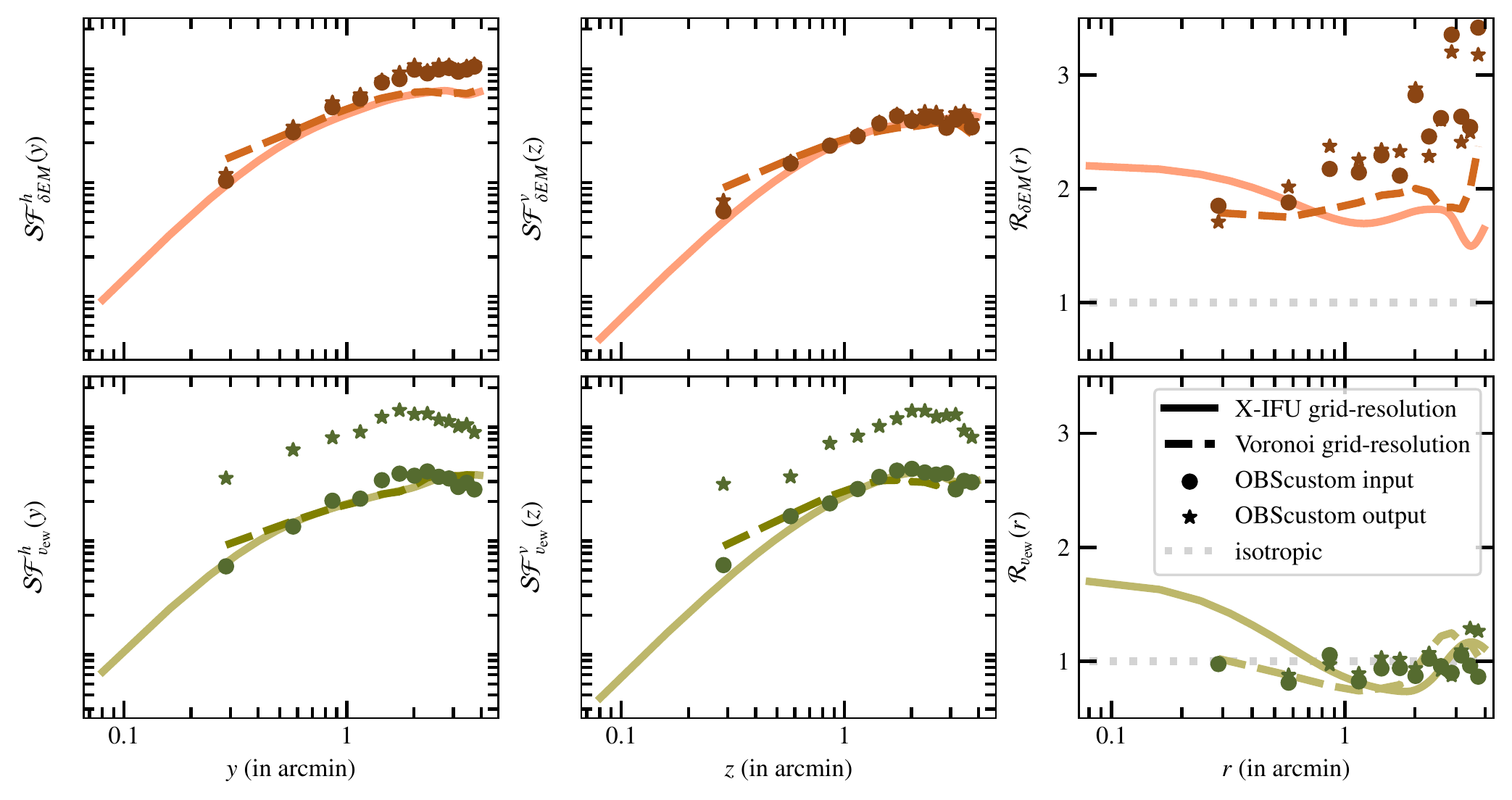}
      \caption{Diagnostic tools for probing the anisotropic structure of turbulent fields.
              From left to right: second-order horizontal and vertical structure functions
              and their ratios
              for the emission measure fluctuations (top) and emissivity-weighted velocity (bottom).
              Full (resp. dashed) lines correspond to the quantities
              computed from the rescaled MHD simulation
              on a X-IFU--like (resp. Voronoi) grid
              with $12$ independent MTI turbulent fields.
              The dots represent the same ratios, but derived from a
              single rescaled field of the MHD simulation used as input to OBScustom
              (top maps in Fig. \ref{fig:obscustom}).
              They are naturally noisier than those computed from the statistics of
              12 independent fields,
              but also more anisotropic in average
              in the case of the emission measure fluctuations.
              Stars show the ratios as derived from the fields reconstructed from OBScustom
              (bottom maps in Fig. \ref{fig:obscustom}).
              The $y$-axis of the leftmost and centre panels showing the structure functions 
              are in arbitrary units but similar for each row,
              so that the horizontal and vertical structure functions can be directly compared.
              }
      \label{fig:anisf}
   \end{figure*}

\subsubsection{Two synthetic observations}
   We virtually point the X-IFU at $\x R_{200}$
   in a mock galaxy cluster with Perseus-like
   thermodynamic profiles
   ($\rho{\sim} \rhoo \ \mathrm{g/cm^{3}}, \ T{\sim} \To \ \mathrm{keV},
   \ H_\rho{\sim}\Hrho \ \mathrm{Mpc}, \ H_T{\sim}\HT \ \mathrm{Mpc}$)
   and at redshift $\zstroke_0{=}0.0179$
   (the corresponding comoving distance is $77$ Mpc).
   At this cosmological redshift, the field of view of the X-IFU 
   is about
   $\mathrm{fov_{X-IFU}}=120$ kpc (see Fig. \ref{fig:lengths}).
   The
   mock observation and field reconstruction
   pipelines are run two times for different input fluctuation fields:
   \begin{itemize}
   \item In the fiducial observation $\mathrm{OBSfid}$,
   we take the fluctuation fields from the 
   Braginskii-MHD simulation
   in the phase of developed turbulence
   at the very end of the simulation, and
   rescale their intensities
   and length scales according to the
   phenomenological discussion from Section \ref{sec:char}, Fig. \ref{fig:3Drms}.
   This translates into $\ell_i{\approx}\li$ kpc (that is $2.6 \mathrm{fov_{X-IFU}}$),
   $\left. \varv_\mathrm{ew} \right|_\mathrm{rms}{\approx}10$ km/s and $\left. \delta T_\mathrm{s}\right|_\mathrm{rms}{\approx}0.5\%$.
   This run will allow us to formally determine the X-IFU detectability limits
   that we introduced in the phenomenological discussion
   from Section \ref{sec:char}, Fig. \ref{fig:3Drms}.
   We see a posteriori that the fluctuation fields reconstructed
   from this synthetic observation are too noisy to derive precise physical diagnostics
   of turbulence.
   This is why we performed a second virtual observation.
   \item
   Theoretical and observational uncertainties related to the estimates from the
   phenomenological discussion from Section \ref{sec:char}
   necessarily remain because of all the assumptions that we made.
   We therefore lead the following formal exercise to better understand what physical information
   could be deduced from a mock X-IFU observation if the fluctuation fields of the MTI
   ended up being detectable:
   we perform a second virtual observation $\mathrm{OBScustom}$,
   in which
   the properties of the fluctuation fields
   are optimally customise.
   We set $\ell_i{\approx}\licustom$ kpc,
   $\left.\varv_\mathrm{ew}\right|_\mathrm{rms}{\approx}50$ km/s
   and
   $\delta T_\mathrm{s}|_\mathrm{rms}{\approx}5\%$,
   which ensure that the fluctuation levels are
   above the X-IFU detectability limits
   identified in the previous run.
   OBScustom corresponds to a kind of best-case scenario in which
   the intensities of the temperature fluctuation and velocity fields
   are respectively increased by factors of $10$ and $5$ with respect
   to the fiducial synthetic observation OBSfid.
   \end{itemize}

   In each case, we simulated an observation of $\texp$ Ms.
   Although such long observations are not currently forecasted with
   the ATHENA/X-IFU, we stand by this value for the needs of this formal
   and prospective exercise. We note that 
   \citet{roncarelli} used similar exposure times.
   2 Ms of observation leads to
   $N_\mathrm{ph} \sim 15 \times 10^6$ photons
   in the $0.2-12$ keV range in both cases. Such a
   total number of photons is quite low
   given the very large exposure time used,
   the reason being that we are looking
   at the cluster periphery
   where the MTI is expected to grow,
   rather than
   its centre as was done, for instance, in
   \citet{roncarelli} and \citet{cucchetti}.
   Given the small number of photons,
   pixels need to be gathered into bins using the Voronoi tessellation method
   \citep{voronoi}
   to obtain satisfying spectral fits.
   We found that grouping pixels into regions with
   ${\sim}\Nphbin$ photons in average was sufficient
   in this respect.
   This leads to ${\sim}\Nbin$ Voronoi regions in total.
   The extracted spectra are then fitted with the APEC model using the
   X-ray fitting package XSPEC.

   \subsubsection{X-IFU maps of integrated and reconstructed quantities}
   To assess the quality of the 2D fields
   inferred from a synthetic observation,
   they should be compared
   to the actual input 3D physical quantities
   integrated along the line of sight.
   We therefore computed
   the numerical counterparts of the physical quantities
   Eqs. (\ref{eq:vew})-(\ref{eq:Ts})-(\ref{eq:EM})
   from the rescaled MHD simulation
   on a 2D X-IFU--like grid divided into Voronoi regions:
\begin{equation}
   \varv_\mathrm{ew} = \left. \sum_i \rho_i^2 \varv_{x,i}\right/\sum_i \rho_i^2,
   \label{eq:vewnum}
\end{equation}
\begin{equation}
   T_\mathrm{s} = \left. \sum_i \rho_i^2 T_i^{0.25}\right/\sum_i \rho_i^2 T_i^{-0.75},
   \label{eq:Tslnum}
\end{equation}
\begin{equation}
   EM = \sum_i \rho_i^2 \ell_x,
   \label{eq:EMnum}
\end{equation}
   where $i$ runs over all cells
   pertaining to a given Voronoi region,
   therefore summing on cells with close enough
   position in the sky
   but possibly very different positions along the line of sight. 
   In the context of subsonic turbulence,
   we are equally interested in the simulated and observed thermodynamic fluctuations
   $\delta T_\mathrm{s}$ and $\delta EM$,
   whose cumbersome derivation is deferred to Appendix \ref{app:fluc}.

   Fig. \ref{fig:obsfid} shows X-IFU Voronoi-tesselated maps
   from the synthetic observation OBSfid, physically motivated
   by Section \ref{sec:char}, Fig. \ref{fig:3Drms}.
   The emission measure fluctuation (leftmost maps)
   and emission-weighted velocity (rightmost maps) fields
   are partially recovered, although they look very noisy with respect to their
   respective input maps.
   In the case of the spectroscopic temperature fluctuations (middle panels),
   the reconstructed signal is overwhelmed
   by the observational noise due to both
   a lack of photons and instrumental limitations.
   From the quantitative analysis of the biases and standard deviations of the reconstructed output fields
   with respect to the true input fields presented in Appendix \ref{app:analysis}, we know that
   the X-IFU detectability limits for a 2-Ms
   observation at $\x R_{200}$ of the Perseus cluster are about
   $\left.\varv_\mathrm{ew}\right|_\mathrm{rms, min}{\sim}20\ \mathrm{km/s}$,
   $\left.\delta T\right|_\mathrm{rms,min}{\sim}2\%$.
   We note that, even on the ideal input maps from the rescaled MHD simulation (top panels),
   very few turbulent structures are seen because the MTI injection length
   at $\x R_{200}$ in a Perseus-like cluster is $\li$ kpc,
   which is $2.6$ times larger than the X-IFU field of view.
   
   Because of both weak fluctuation intensities and large injection length,
   OBSfid cannot be used to satisfactorily derive
   usual physical diagnostics like turbulent spectra or structure functions.
   We therefore performed the synthetic observation $\mathrm{OBScustom}$
   with optimally customised parameters
   to further check our procedures for building mock observation
   and for reconstructing physical fluctuation fields.
   We stress that these parameters are not directly motivated by \citetalias{perrone2D}
   scaling laws.
   They will however allow us
   to estimate to which extent the anisotropic structure of MTI turbulence
   may be constrained with the perturbation fields available from a X-IFU observation.
   Fig. \ref{fig:obscustom} shows that the fluctuation fields reconstructed from OBScustom
   (for which the root-mean square of the input fluctuation fields
   now exceed the expected observational noise)
   are visually well recovered.

   Altogether, we now know to which extent the capabilities
   of the X-IFU would allow us to retrieve, or not, thermodynamic and velocity fluctuations
   according to their expected intensities.
   More specifically, the X-IFU detectability thresholds are
   $\left.\varv_\mathrm{ew}\right|_\mathrm{rms, min}{\sim}20\ \mathrm{km/s}$,
   $\left.\delta T\right|_\mathrm{rms,min}{\sim}2\%$
   for a 2-Ms observation of a Perseus-like cluster at $\x R_{200}$.
   In the next section, we assess whether a signature of MTI turbulence,
   namely its anisotropic structure with respect to the direction of the temperature gradient,
   could be observationally detected
   if the fluctuation intensities
   turned out to be higher
   (by factors of 5 to 10)
   than the physical estimates used to scale OBSfid.
   For this purpose, we use the synthetic observation OBScustom.

   \subsection{Anisotropic structure functions}
   We now investigate the detectability of MTI-induced anisotropy
   in the best-case scenario, represented by the synthetic observation OBScustom.
   To this end,
   we introduce the horizontal and vertical second-order structure functions and their ratio:
\begin{equation}
\label{eq:sfh}
   \mathcal{SF}_X^h(y) = <\left[X\left(\vec{c} + y\vec{e_y}\right) - X\left(\vec{c}\right)\right]^2>,
\end{equation}
\begin{equation}
\label{eq:sfv}
   \mathcal{SF}_X^v(z) = <\left[X\left(\vec{c} + z\vec{e_z}\right) - X\left(\vec{c}\right)\right]^2>,
\end{equation}
\begin{equation}
   \mathcal{R}_X (r) = \left.\mathcal{SF}_X^h(r)\right/\mathcal{SF}_X^v(r),
\end{equation}
   where $X$ is either $\delta EM$ or $\varv_\mathrm{ew}$,
   the spatial average runs over the centres $\vec{c}$ of the Voronoi regions.
   We set $y{=}z{=}r$ in the last equation.
   We ignore the spectroscopic temperature fluctuations as, in our setup,
   the physical information they contain is
   redundant with, and of worse quality than,
   that found in the emission measure fluctuations.
   In isotropic turbulence,
   scalar structure functions does not depend
   on the particular direction along which they are computed
   and the quantity $\mathcal{R}_X$ is thus equal to unity
   at all scales
   (statistically speaking at least)
   for any isotropic scalar field $X$.
   This ratio provides somehow a measurement of the anisotropic structure
   of a turbulent field \citep{rincon06}.
   $\mathcal{R}_X>1$ indicates that the vertical injection length $\ell_v$
   is greater than the horizontal injection length $\ell_h$
   (corresponding to vertically elongated structures)
   because the second-order horizontal
   and vertical structure functions
   $\mathcal{SF}_X^{h,v}$
   are expected to tend
   towards $0$ for $y,z{\ll}\ell_h,\ell_v$,
   towards the same value ${<}X^2{>}$ when $y,z{\gg}\ell_h,\ell_v$,
   and to increase in-between.

   This new diagnostic is used to characterise the anisotropy
   of the emission measure fluctuation and emissivity-weighted velocity fields
   from the best-case synthetic observation OBScustom.
   All relevant quantities related to this diagnostic are presented in Fig. \ref{fig:anisf}.
   Dots and stars on the rightmost panels
   show the ratio $\mathcal{R}_X$ as derived
   from the input fields of this mock observation
   (top maps in Fig. \ref{fig:obscustom}),
   and from the reconstructed fields
   (bottom maps in Fig. \ref{fig:obscustom}),
   on the corresponding Voronoi grid;
   whereas the horizontal and vertical structure functions
   $\mathcal{SF}_X^{h,v}$
   are plotted in the leftmost and centre panels respectively.
   In all cases but the structure functions of the emission-weighted velocity field,
   data computed from the input fields of the virtual observation (dots) 
   and from solving the inverse problem (stars)
   are in broad agreement.
   This is actually expected
   when the statistical and systematic errors
   (gathered into the dispersion $\sigma_X$ in this work)
   are small enough with respect to the quantity $X$ itself,
   as previously demonstrated by \citet{zuhone}, \citet{roncarelli}, and \citet{cucchetti}
   for isotropic second-order structure functions.
   In the case of the reconstructed emissivity-weighted velocity field from OBScustom,
   the statistical dispersion $\sigma_{\varv_\mathrm{ew}}$
   with respect to the true input field
   is of the order of the root-mean square of the true input field.
   The structure functions computed from the reconstructed emissivity-weighted velocity field
   (green stars on the leftmost and centre bottom plots in Fig. \ref{fig:anisf})
   are therefore uniformly biased with an additional $\sigma^2_{\varv_\mathrm{ew}}$ term
   with respect to the structure functions computed from the true input field
   (green dots).
   This effect is independent of the direction in which the structure function is computed,
   so it does not affect the ratio of the structure functions itself.

   We overall see that the emission measure fluctuation field
   is relatively anisotropic $\mathcal{R}_{\delta EM}{\gtrsim}2$
   while the emission-weighted velocity field is rather isotropic
   $\mathcal{R}_{\delta \varv_\mathrm{ew}}{\sim}1$.
   To check that these specific features are robust against
   the variance associated with a single realisation of a MTI turbulent field,
   we accumulated better statistics
   by computing the same ratios 
   and structure functions
   (dashed lines in Fig. \ref{fig:anisf}),
   on the same Voronoi grid,
   from $3$ different snapshots of the simulation at saturation,
   separated by at least one turnover time.
   Each snapshot represents,
   after integration along the line of sight,
   $4$ independent realisations of a MTI turbulent field
   since $\ell_i/L{\sim}22\%$ at saturation.
   We still get $\mathcal{R}_{\delta EM}{\sim}2$.
   To further check that these specific features does not originate from
   binning X-IFU pixels into bigger Voronoi regions,
   we also computed
   $\mathcal{R}_X$ and
   $\mathcal{SF}_X^{h,v}$
   on the original X-IFU grid
   (full lines in Fig. \ref{fig:anisf}).
   In this case, the fields from the simulation
   are first degraded at X-IFU pixel resolution
   and the average in Eqs. (\ref{eq:sfh})-(\ref{eq:sfv}) runs over the centres of X-IFU pixels.
   Binning pixels into Voronoi regions does not alter
   the degree of anisotropy seen in the fields,
   although it slightly changes the shape of the ratio $\mathcal{R}_X$.

   We are therefore able to reliably identify MTI-induced anisotropy
   in the emission measure fluctuation field
   with an ideal
   2-Ms observation of a Perseus-like cluster at $\x R_{200}$,
   if the actual levels of MTI turbulent fluctuations
   at saturation were to be five to ten times higher
   than the current theoretical estimates from Section \ref{sec:char}, Fig. \ref{fig:3Drms}.
   Finally, we note that our technique is broadly applicable to the
   detection of anisotropic dynamics beyond the case of the MTI.

\section{Conclusions and discussion}
\label{sec:discussion}

\subsection{Summary and main conclusions}
\label{sec:conclusion}

   The current work was intended as a first step
   towards bridging
   the gap between theoretical astrophysical plasma studies
   and future X-ray observations with the X-IFU
   on board the future European X-ray observatory ATHENA,
   given the instrumental specifications targeted before the recent design-to-cost exercise
   (the discussion is extended to present X-ray telescopes later in this section).
   We characterised the dynamical properties of the MTI,
   a magnetised buoyancy instability
   suspected to operate at fluid scales in the ICM
   (as opposed to micro-scale kinetic instabilities)
   and conducive to the anisotropic conductive heat flux,
   and we assessed whether it would be detectable with the X-IFU.

   Our main conclusion comes from the combination of the results of
   Sections \ref{sec:char} and \ref{sec:results}:
   while the raw physical estimates for MTI fluctuations
   suggest that they should be detectable, we have shown that several physical
   and observational factors conspire in a non-trivial, cumulative way
   to put the actual observational
   MTI signal close to the X-IFU detectability limits.
   More specifically,
   fluctuation fields from a Braginskii-MHD simulation,
   rescaled in a physically motivated way
   according to the results of Section \ref{sec:char}
   for a Perseus-like cluster at $\x R_{200}$,
   and used as inputs of the
   2-Ms synthetic observation OBSfid
   cannot be satisfactorily recovered
   (but only marginally so and this was not obvious in the first place).
   Yet this virtual observation 
   corresponds in many ways to an ideal scenario because of
   the very large exposure time, as well as the absence of background noise
   and of any other sources of turbulence but the MTI.
   Our analysis suggests that
   MTI-induced turbulence,
   if it is indeed described by the physical scaling laws of \citetalias{perrone2D} in cluster haloes,
   would not be detectable in the real ICM,
   even with the upcoming generation of X-ray spectrometers such as the X-IFU.
   We identified several effects conspiring
   to hamper the detectability of MTI turbulence
   in ICM regions amenable to X-ray observations:
\begin{itemize}
\item The levels of MTI turbulence at saturation, as computed from
      the scaling laws derived by \citetalias{perrone2D}, are only moderate
      for radii between $0.2$ and $0.6R_{200}$
      with $\left.\varv\right|_\mathrm{rms}{=}\mathcal{O}(100\ \mathrm{km/s})$,
      $\left.\delta T\right|_\mathrm{rms}{=}\mathcal{O}(5\%)$
      for the three different ICM models of thermodynamic profiles.
\item The contribution of the azimuthal (i.e. horizontal) component over the radial (i.e. vertical)
      component of the velocity field is predominant
      on the observable line-of-sight velocity in the halo of galaxy clusters,
      where the MTI can develop.
      The former is significantly smaller than the latter
      because of the inherent anisotropy of MTI turbulence
      with respect to the direction of gravity,
      the root-mean square of the line-of-sight velocity field
      is then
      $\left.\varv_\mathrm{los}\right|_\mathrm{rms}{=}\mathcal{O}(50\ \mathrm{km/s})$.
      This reduction is stronger than in the case of isotropic turbulence. 
\item The root-mean square of the measurable turbulent fluctuations are also lowered
      because of the statistical cancellation of opposite-sign fluctuations
      present along the line of sight
      due to the expected MTI injection scale (${\sim}300$ kpc) being smaller than the
      typical length over which significant emission takes place (${\gtrsim}$Mpc).
      This effect, which has been statistically estimated thanks to Eq. (5) from \citet{clerc},
      leads to
      $\left.\varv_\mathrm{ew}\right|_\mathrm{rms}{=}\mathcal{O}(10\ \mathrm{km/s})$,
      $\left.\delta T\right|_\mathrm{rms}{=}\mathcal{O}(1\%)$.
      It is likely to intrinsically limit the ability of current and future X-ray observations
      to probe small-scale fluctuations of MTI and other types of turbulence,
      and to measure small-scale spectral energy density below some critical threshold.
\item Finally, there is no robust cross-correlation between
      the horizontal velocity and the density fluctuations in MTI-induced turbulence,
      which could otherwise have been exploited if present.
\end{itemize}

   We made some assumptions and used some simplifications when we
   phenomenologically quantified the observable dynamical properties of the MTI:
   uncertainties related to the previous estimates necessarily remain.
   An additional synthetic observation OBScustom was therefore performed to anticipate to what extent
   the conclusions would differ if the estimations of the observable MTI turbulent levels,
   as expected from Section \ref{sec:char} and Fig. \ref{fig:3Drms}, are too weak in reality.
   We have shown that,
   when the intensity of the fluctuations is higher than the X-IFU detectability limits,
   the inherent vertically anisotropic structure of MTI-driven turbulence can be detected
   in the emission measure fluctuation field as recovered
   from a 2-Ms observation with the X-IFU.
   Obviously though,
   this characteristic
   is neither a strong observational signature of, nor specific to, the MTI
   since many other processes could drive anisotropic turbulence in the ICM.

   More broadly,
   our study confirms that directly identifying plasma effects
   in the internal magnetised dynamics of the ICM
   remains a difficult observational task
   given the intrinsic performance of X-ray instrumentation.
   We further discuss the possible detection of anisotropic MTI dynamics
   at large fluid scales
   in \ref{sec:strategy}.
   Such an enterprise may benefit from the advent of large radio telescopes
   such as the SKA to combine observations of the ICM dynamics with observations
   of its magnetic geometry, enabling the use of cross-diagnostics to identify
   magnetised plasma structures in the ICM.
   
\subsection{Validity of the scaling laws and magnetic feedback}
\label{sec:validity}

   The phenomenological estimates of the measurable length scales and intensities of
   MTI turbulent fluctuations
   depicted in Section \ref{sec:char}
   are subservient to the validity
   of the scaling laws derived in \citetalias{perrone2D}.
   Though we do believe that these relations hold in regions of galaxy clusters
   amenable to X-ray observations,
   on dimensional grounds the scaling for the buoyancy potential energy
   Eq. (\ref{eq:Upropto})
   is currently incomplete and might be subject to revision
   before the root-mean square of the density and temperature fluctuations
   can be self-consistently computed.
   Furthermore, additional work would be required to determine whether these scaling laws
   are robust against the development of kinetic instabilities triggered by pressure anisotropy.
   For instance, a well-known feature of electron-scale
   instabilities
   is the reduction and/or isotropisation of the heat flux \citep{riquelme,komarov,drake},
   which would lower the MTI injection scale according to Eq. (\ref{eq:li}).
   The intensity of the MTI thermodynamic and velocity fluctuations
   would also suffer from such a reduction of the Spitzer conductivity,
   if we assume Eqs. (\ref{eq:K})-(\ref{eq:U})
   to hold as such in the presence of micro-scale kinetic instabilities,
   thereby further compromising the detection of any relevant MTI signal
   from an X-IFU observation.

   At this stage, it is also relevant to wonder whether (and, if so, how)
   the structure of MTI turbulence depends on the feedback of the magnetic field
   on the flow through the Lorentz force when equipartition between magnetic and kinetic energies is reached.
   This is indeed a difference between the fiducial simulation of \citetalias{perrone3D}
   and ours
   because of the higher viscosity and resistivity we used in our simulation,
   and which might bias our conclusions
   on the possible anisotropy of the relevant fluctuation fields.
   From the \citetalias{perrone3D} fiducial simulation, it seems that the magnetic feedback tends
   to make the flow even more anisotropic at equipartition.
   This peculiarity may,
   at the same time,
   lower the intensity of the observable line-of-sight velocity
   in the outskirts of the ICM,
   and exacerbate the anisotropy of the density fluctuations,
   which we were already able to detect with an ideal X-IFU synthetic observation
   of the Braginskii-MHD simulation yet not at equipartition.
   In any case, 
   determining whether the MTI drives a dynamo in the ICM
   requires a clear picture of the ICM effective collisionallity.
   Such a picture is still currently missing
   but is under intense theoretical and numerical investigation
   \citep{rincon16,stonge2018,stonge2020,squire2019,squire2023,arzamasskiy2023}.

   As a final note,
   we emphasise that the vertical and horizontal injection lengths
   of MTI-induced turbulence may be subject to different scaling laws,
   although we assumed Eq. (\ref{eq:li}) for both.
   This could also bias the degree of anisotropy
   of realistic MTI turbulent fields, with respect to those used in this study.

\subsection{Future observational strategies}
\label{sec:strategy}

   We found that directly detecting features of MTI turbulence
   in the ICM would be very demanding, even with the X-IFU.
   Yet what we learnt from trying to do so and what this implies for future observations
   are worth discussing.
   A possible observational strategy for the detection of MTI-driven turbulence
   could be to aim the instrument at the Perseus cluster and build a mosaic of X-IFU maps
   with a shorter exposure time for each observation.
   Given the large injection length expected for MTI-driven turbulence,
   ${\sim}34$ Voronoi regions (instead of ${\sim}340$ currently)
   seem enough to satisfactorily map the X-IFU field of view
   without degrading the spatial resolution of the observed field too much.
   Assuming the same cumulated exposure time
   and signal-to-noise ratio requirement,
   ten observations of $200$ ks each would roughly lead to the same total number
   of Voronoi regions with an average global effective field of view about
   three times that of X-IFU. The ratio between MTI injection length
   and such a global field of view would then be about ${\sim}0.9$,
   which is quite close to the ratio of $0.7$ used in the virtual observation $\mathrm{OSBcustom}$,
   partially justifying the approach adopted
   for prospective purposes
   in the last synthetic observation,
   at least in terms of the injection scale
   (providing similar arguments to justify the boost of the velocity and 
   thermodynamic fluctuations is more challenging, as discussed in \ref{sec:validity}).
   In addition, this strategy would offer some flexibility
   in the choice of the pointing configuration,
   which could then be optimised
   for the detection of MTI turbulent features,
   as explored by \citet{zuhone} with Hitomi in the case of isotropic turbulence.

   We have shown that
   the MTI-driven anisotropy is better inscribed 
   in the emission measure (i.e. density) fluctuations
   than in the emissivity-weighted velocity field,
   in the periphery of galaxy clusters at least.
   The detection of such density perturbations does not require
   X-ray spectrometers.
   However, we argue that current
   XMM-Newton, Chandra, and XRISM observatories
   would not collect enough photons with a reasonable exposure time
   to probe the anisotropic dynamics of the faint cluster haloes.
   The next generation of X-ray telescopes such as the ATHENA/X-IFU will increase
   the photon collecting area by a factor ${\sim}10$ relative to current
   X-ray missions.
   They will therefore provide
   key observations of dim ICM outskirts, precisely in those regions most
   favourable to the development of the MTI.

\subsection{Competing sources of anisotropic turbulence in the ICM}
\label{sec:competing}

   The MTI is not the only source of anisotropic turbulence in the ICM.
   We now discuss the role of stable entropy stratification and accretion in this context.

   Using idealised high-resolution simulations
   of compressible or solenoidal stratified turbulence,
   \citet{mohapatra20,mohapatra21}
   were able to numerically relate the ratio between
   the root-mean square of
   the vertical $\left.\varv_z\right|_\mathrm{rms}$
   and the horizontal $\left.\varv_h\right|_\mathrm{rms}$ velocity
   with the Froude number.
   The latter is
   defined as $\mathrm{Fr}=\left.\varv\right|_\mathrm{rms}/(\ell_i N)$
   and compares the timescales associated with the turbulence and with the stratification.
   Injecting the scaling laws from \citetalias{perrone2D}, we find
   $\mathrm{Fr} \approx 0.15 \omega_T/N$
   for MTI-driven turbulence, which is about ${\sim}0.1$
   in regions of galaxy clusters amenable to X-IFU observations.
   This estimation lies right in the range that
   \citet{mohapatra20,mohapatra21} identified as the more
   susceptible to the effects of stratification.
   In their simulations,
   the turbulent forcing is however isotropic
   (or, in some cases, perpendicular to the direction of gravity):
   MTI-induced turbulence
   exhibits a different structure and
   may not be subject to the exact same physics
   in the presence of strong entropy stratification.
   At this stage, it is still unclear whether
   the stable stratification
   will isotropise MTI turbulence,
   in the stratification regime relevant to the ICM $(\mathrm{Fr}{\sim}0.1)$,
   or whether
   it will stay vertically anisotropic,
   on account of its convective nature.
   Some clues can however be gained from our simulation and from
   the most stratified MTI run in \citetalias{perrone3D},
   which used
   $N{\approx}0.8\omega_T,\ \mathrm{Fr}{\sim}0.2$
   and $N{=}\omega_T, \ \mathrm{Fr}{\sim}0.15$, respectively,
   and are therefore not too far from real ICM stratification regimes.
   These simulations suggest that the MTI vertical anisotropy remains active despite the
   strong entropy stratification.
   Further progress on this question would require additional
   high-resolution numerical simulations
   of strongly stratified MTI turbulence, in the regime relevant to the ICM.
   
   On the other hand, \citet{simonte} and \citet{vazza} argue that
   local simulations of stratified turbulence are too idealised
   and that, in real clusters, global accretion drives, preferentially, radial turbulence
   with enough momentum and energy
   to overcome the effect of the stable stratification.
   Accretion could thus be another relevant source
   of vertical anisotropy in the ICM.
   However, with this turbulent forcing being compressible with Mach numbers $\gtrsim0.3$,
   the velocity field is no longer divergence-free and
   the anisotropy between the radial and azimuthal kinetic energies
   may reflect less on the structure of the turbulent eddies.

   High-resolution global simulations including MTI and other magnetised plasma effects,
   internal stirring mechanisms, and accretion-driven
   turbulence are needed to analyse, more precisely, the competition
   between different sources of turbulence in the ICM
   and to draw a clearer picture of the ICM internal dynamics
   in the outskirts of galaxy clusters.
   Such global MHD simulations could also be virtually observed through the X-IFU,
   and their internal dynamics reconstructed, to assess what turbulent components
   dominate and could be identified with a mock observation.
   Previous larger-scale cosmological style simulations
   with anisotropic heat conduction
   failed to shed light on
   any distinctive MTI flow in cluster peripheries \citep{ruszkowski}.
   \citetalias{perrone2D} however clearly showed that MTI saturation is very dependent
   on the level of parallel thermal diffusivity used. Care should therefore be taken to simulate
   the right anisotropic diffusive regime
   and to control perpendicular numerical dissipation as much as possible
   if one wants to detect any MTI flow in such simulations of structure formation.

\begin{acknowledgements}
   The authors are grateful to CNES for financial support provided by Didier Barret
   for the internships of J. M. K. during which a major part of this work was conducted. 
   J. M. K. is thankful for the organisation of the SIXTE workshops
   by the development team, allowing him a smooth start with the software.
   The work benefited from very fruitful exchanges
   with Lorenzo Perrone, Henrik Latter,
   Simon Dupourqué and Alexeï Molin.
   The authors also thank Geoffroy Lesur for
   giving them access to a beta version of IDEFIX
   and for his support in the integration
   of the Braginskii physical modules.
   This work was granted access to the HPC resources of CALMIP
   under allocation 2021/2022-P16006 and the authors thank the associated support team.
   We are finally grateful for the generous hospitality of the Wolfgang Pauli Institute, Vienna, where
   preliminary results were discussed during the 2022 Plasma Kinetics workshop.
\end{acknowledgements}

%
%

\bibliographystyle{aa}
\bibliography{ref}

\begin{appendix} 
\section{Thermodynamic profiles}
\label{app:prof}

    All thermodynamic profiles
    presented in Section \ref{sec:method}
    and used to derive
    the properties of MTI-induced turbulence
    in Section \ref{sec:char}
    can be parametrised as \citep{vikhlinin,ghirardini}:
\begin{equation}
    \frac{T(x)}{T_{500}} = T_1 \frac{\frac{T_{\mathrm{min}}}{T_1}+\left(\frac{x}{r_{\mathrm{cool}}}\right)^{a_{\mathrm{cool}}}}{1+\left(\frac{x}{r_{\mathrm{cool}}}\right)^{a_\mathrm{cool}}}\frac{1}{\left(1+\left(\frac{x}{r_t}\right)^2\right)^{\frac{d}{2}}},
\end{equation}
\begin{equation}
    \rho(x) = \rho_1 \frac{\left(\frac{x}{r_\mathrm{c}}\right)^{-\frac{\alpha}{2}}}{\left(1+\frac{x^2}{r^2_\mathrm{c}}\right)^{\frac{3\beta}{2} - \frac{\alpha}{4}}}\frac{1}{\left(1 + \frac{x^\theta}{r_\mathrm{s}^\theta} \right)^{\frac{\epsilon}{2\theta}}},
\end{equation}
   with $x=r/R_{500}$, $T_{500} = 5$ keV and
   where we simply set $R_{500}{=}1.1$ Mpc for all clusters.
   All the other quantities are parameters
   to fit.
   For the universal thermodynamic profiles from \citet{ghirardini},
   the best-fit parameter values from that paper
   for CC clusters are directly used.
   Fits are performed for Perseus temperature and density profiles \citep{simionescu},
   but also for the 1D-spherical accretion model of \citet{mccourt}
   since the data resulting from the numerical integration of this model are not
   perfectly robust against numerical derivation
   (which we need to obtain the local scale-height of the different atmospheres).
   In the case of the density, the fit is not performed on the data but rather
   on their logarithm, so that the weak tail of the density profiles are well fitted too.
   The best parameter values are given in Table \ref{tab:prof} for the different profiles.
   We do not fit the data from the different entropy profiles.
   Instead, we assume perfect hydrostatic equilibrium to deduce
   the entropy scale-height from the temperature and density
   scale-heights:
\begin{equation}
    H_S = \frac{H_\rho H_T}{H_\rho - \left( \gamma -1 \right) H_T}.
\end{equation}

\begin{table*}
\caption{\label{tapp}Set of parameters for the
temperature and density profiles of the different thermodynamic models of ICM.}
\centering
\begin{tabular}{lcccccc}
\hline\hline
Temperature model & $T_1$ & $T_\mathrm{min}$ & $r_\mathrm{cool}$ & $a_\mathrm{cool}$ & $r_\mathrm{t}$ & $d$ \\
\hline
\citet{ghirardini} & 1.3 & 0.29 & 0.061 & 0.74 & 0.40 & 0.66 \\
\citet{mccourt}    & 0.38& 0.49 & 0.47 & 3.0 & 2.7 & 9.5 \\
\citet{simionescu} & 1.4 & 0.00 & 0.12 & 10 & 1.3 & 1.7 \\
\hline
\hline
\end{tabular}
\begin{tabular}{lcccccc}
Density model & $\rho_1\left(\mathrm{in}\ 10^{-27}\mathrm{g/cm^3}\right)$ & $r_\mathrm{c}$ & $\alpha$ & $\beta$ & $r_\mathrm{s}$ & $\epsilon$ \\
\hline
\citet{ghirardini} & 0.020 & 0.041 & 0.80 & 0.49 & 1.2 & 4.7 \\
\citet{mccourt}    & 22 & 0.26 & 3.4 & 0.83 & 2.0 & 20 \\
\citet{simionescu} & 7.3 & 0.00051 & 0 & 0.90 & 1.4 & 11 \\
\hline
\end{tabular}
\tablefoot{
We enforced $\theta = 3$ in all cases
and $\alpha=0$ in the case of Perseus
for convergence purposes \citep{shi,dupourque}.
}
\label{tab:prof}
\end{table*}

\section{Validation of the Braginskii operators in IDEFIX}
\label{app:brag}
   We implemented the parabolic operator for anisotropic heat diffusion
   Eq. (\ref{eq:qbrag}) with the centred asymmetric scheme described
   in \citet[Section 2.1]{sharmahammet},
   though without harmonic mean nor slope limiter.
   The Braginskii viscosity Eq. (\ref{eq:pibrag}) is implemented 
   in a similar way, again without any slope limiter. 
   The implementation of these operators in IDEFIX is validated thanks to the
   methodology and setups described in \citet[Sections 3,4]{parrish2012b}.
   The linear growth rates of both MTI and HBI diagonal modes
   that we numerically measured are compared with the theoretical values
   expected from the dispersion relation.

\begin{figure}
   \centering
   \includegraphics[width=\hsize]{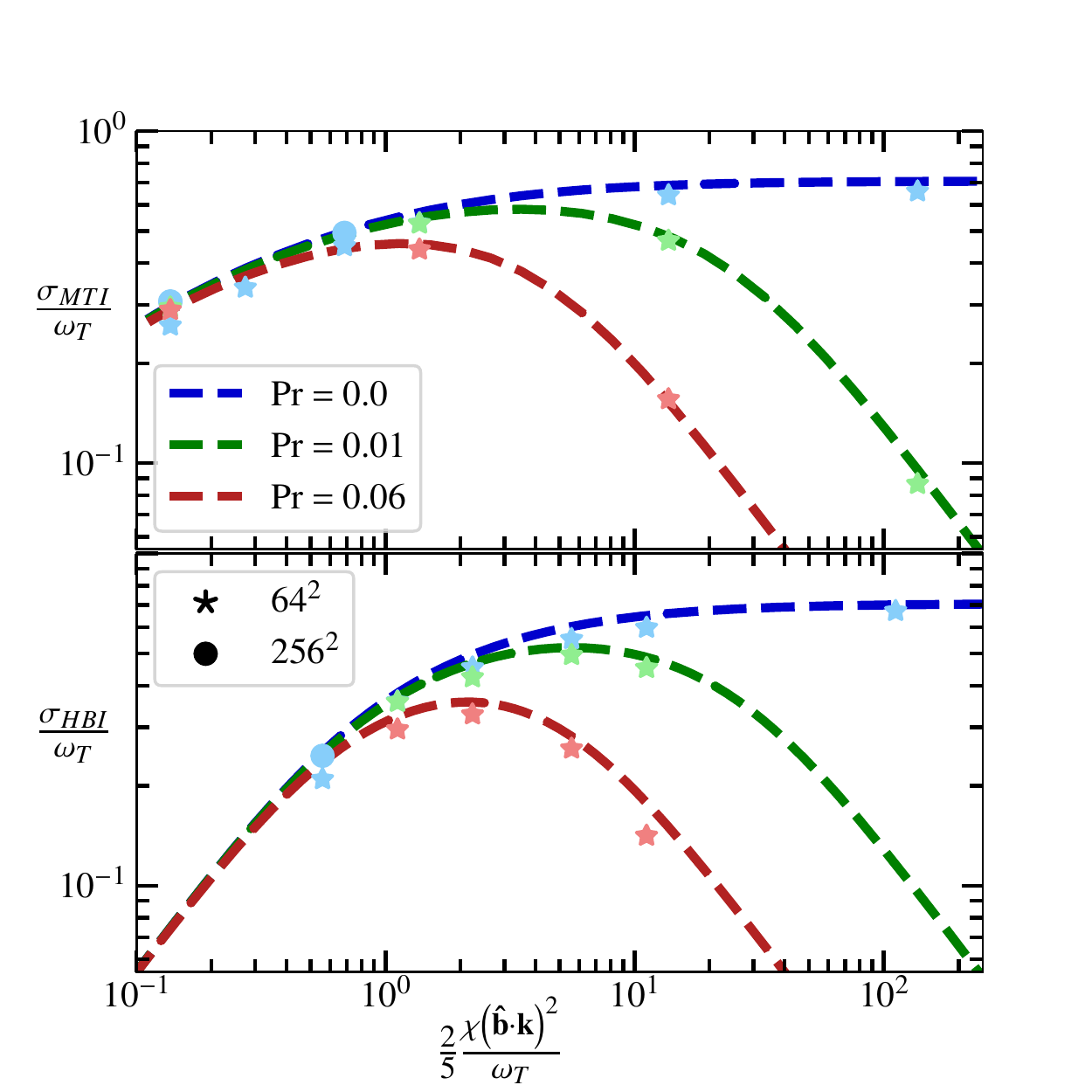}
      \caption{
              Comparison between the MTI (top) and HBI (bottom) linear growth rates numerically
              measured and theoretically computed thanks to the dispersion relations.
              The initial velocity perturbation is
              a fundamental divergence-free 
              diagonal mode $(k_y=k_z=2\pi/L)$ with amplitude $10^{-4}$.
              Different colour tones are for the different Prandtl numbers.
              Runs with resolution $64^2$ and $256^2$ are represented by stars and dots.
              }
      \label{fig:check}
   \end{figure}

   For the MTI setup, the tests are run in 2D ($\vec{e_y}, \vec{e_z}$) at either low 
   or high resolution ($64^2$ and $256^2$,
   respectively the stars and the dots in Fig. \ref{fig:check})
   in a square box of size $L=0.1$.
   Lengths are normalised by the third of the temperature scale-height,
   velocities and magnetic fields respectively by the thermal velocity
   $\varv_\mathrm{th,0}$ at the bottom of the atmosphere
   and by $\sqrt{\rho_0}\varv_\mathrm{th,0}$.
   The atmosphere is initialised according to the hydrostatic
   equilibrium described by Eqs. (\ref{eq:Tz}-\ref{eq:rhoz})
   with $\alpha=1$ and $H_T{=}3$.
   We initially seed both components of the velocity field with
   a fundamental divergence-free diagonal eigenmode
   $\propto \exp \left( \sigma t + i\left[k_y y+ k_z z\right]\right)$
   with an amplitude of $10^{-4}$ and
   where $k_y=k_z=2\pi/L$.
   The kinematic viscosity $\nu$ is based on the 
   thermal diffusivity through the Prandtl number
   $\mathrm{Pr}=\left\{0;0.01;0.06\right\}$ (respectively the blue, green and red
   curves and points in Fig. \ref{fig:check}).

   The runs are initialised with a purely
   horizontal magnetic field $B_{y,0} = 10^{-5}$.
   In this configuration, the dispersion relation is:
\begin{equation*}
   \sigma \left(\sigma+\frac25 \chi k_y^2 \right)
   \left[ \sigma + 3 \nu k_y^2 \left( 1 - \frac{k_y^2}{k^2} \right) \right]
   + \sigma N^2 \frac{k_y^2}{k^2} + \frac25 \chi \omega_T^2 k_y^2 \frac{k_y^2}{k^2}
   = 0.
\label{eq:dispMTI}
\end{equation*}
   Regarding the HBI setup, the runs are performed in a
   box with $L=0.2$, $H_T=2$ and with the same
   hydrostatic equilibria as \citet{parrish2012b}
   in the corresponding local HBI runs (Section 4.1 in that paper).
   The magnetic field is initialised vertically with $B_{z,0} = 10^{-5}$
   while the initial velocity field is the same than in the MTI runs.
   The dispersion relation is then:
\begin{equation*}
   \sigma \left(\sigma+\frac25 \chi k_z^2 \right)
   \left[ \sigma + 3 \nu k_z^2 \left( 1 - \frac{k_z^2}{k^2} \right) \right]
   + \sigma N^2 \frac{k_y^2}{k^2} - \frac25 \chi \omega_T^2 k_z^2 \frac{k_y^2}{k^2}
   = 0.
\label{eq:dispHBI}
\end{equation*}

   We successfully tried other 2D configurations with higher-order harmonic modes
   as well as 3D setups with the gravity set artificially
   along the $\vec{e_x}$- or $\vec{e_y}$-axis
   to check that the operators are correctly implemented
   in all directions.

\section{Building X-IFU mock observations}
\label{app:building}
   In this Appendix, we compile technical details about the
   post-processing observational pipeline.
   The first subsection is specific to our MHD simulation
   since it describes how to convert its outputs
   to a $(x,y,z)$-spatial representation of the density, temperature and velocity fields
   suited to mock observations.
   The second and third subsections are more generic in the sense that they indicate
   how to solve the forward and inverse problems,
   once the spatial representation of the different fields has been built.

\subsection{Designing boxes adapted to synthetic observations}
\label{app:design}
   In this section, we elaborate on how to build, from the Braginskii-MHD simulation,
   a spatial $(x,y,z)$-representation of the density $\rho$, temperature $T$
   and velocity $\varv_\mathrm{los}$
   adapted to a mock observation through the X-IFU with SIXTE.
   First, the physical box size $L$ is set by matching the numerical injection length in the simulation 
   with the chosen dimensional integral length
   of MTI-driven turbulence,
   which is either picked from Eq. (\ref{eq:li}) for OBSfid
   or customised for OBScustom.
   In the same manner, the fluctuation amplitudes from the MHD simulation at $t{=}\tend t_0$
   are rescaled in such a way that their root-mean square
   match the requested values,
   deduced from Section \ref{sec:char}, Fig. \ref{fig:3Drms}
   for OBSfid
   or again custom-picked for OBScustom.

\begin{figure*}
   \centering
   \includegraphics[width=0.9\hsize]{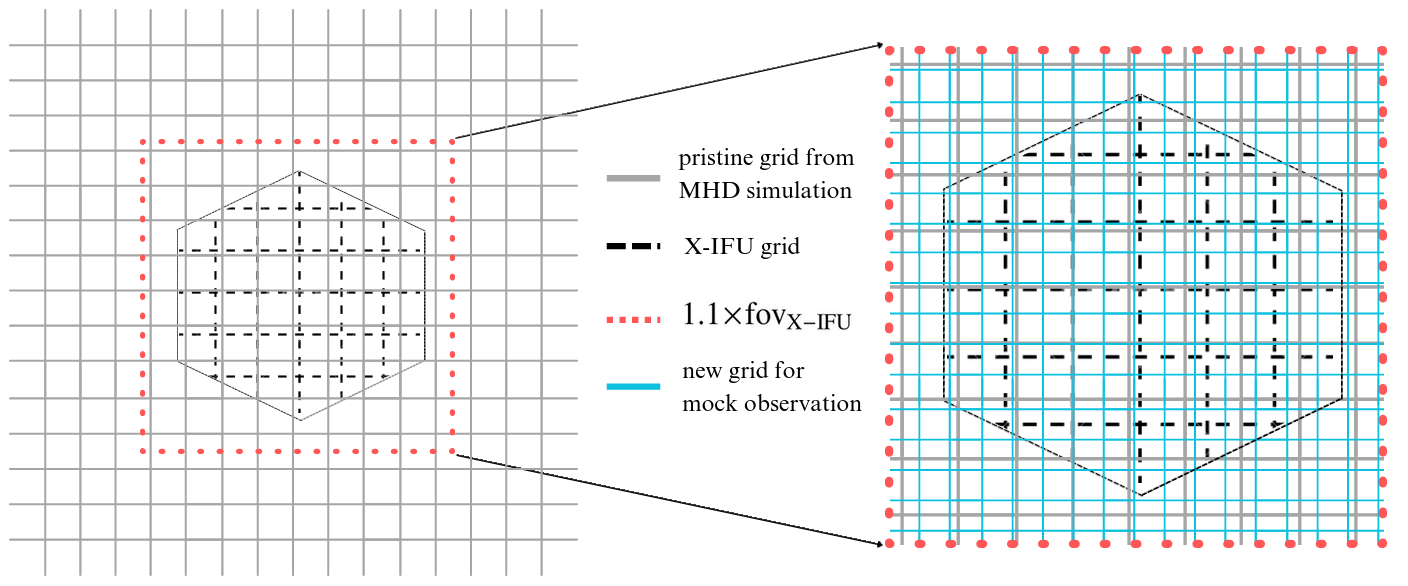}
      \caption{
              Schematic of the different grids used
              in the observational pipeline.
              The relative sizes of the grids are consistent with respect to each other:
              there is almost one cell from the MHD simulation grid (full grey lines, left),
              after rescaling of the box as in OBScustom,
              in a X-IFU pixel (black dashed lines, left)
              and there are ${\sim}4$ cells of the finer grid (light blue lines, right),
              refined from the MHD simulation grid within the dotted red square
              and used for the synthetic observation, in a pixel of X-IFU.
              For the sake of clarity, the absolute sizes of the grids are however not to scale.
              For instance, there should be 3168 pixels in X-IFU's hexagon
              (in our baseline instrumental configuration)
              instead of ${\sim}30$ on the schema, in which X-IFU border pixels are not squared
              whereas they should be as well in reality.
              }
      \label{fig:schematic}
   \end{figure*}

   Next, we extract the fluctuation fields of the rescaled MHD simulation
   in the parallelepiped rectangle extending over the whole box along
   the line of sight (i.e. the $\vec{e_x}$-axis),
   centred on the centre of the box in the $(\vec{e_y}, \vec{e_z})$ plane
   and extending over $1.1 \times \mathrm{fov_{X-IFU}}$ at least 
   (about 130 kpc at $\zstroke_0=0.0179$)
   in this plane.
   In practice, this translates into a parallelepiped rectangle
   with $256\times n^2$ cells from the MHD simulation,
   where $n=\lceil 256 \times 1.1 \times \mathrm{fov_{X-IFU}}/L\rceil$
   is the number of extracted cells
   in one direction $\vec{e_y}$ or $\vec{e_z}$ of the sky plane.
   But virtually observing an ICM-like resolved and extended X-ray source with SIXTE
   requires that the field of view of a single observing pixel contains
   enough individual source points. 
   For instance, in the work of \citet{roncarelli}, one X-IFU pixel
   encompassed ${\sim}8$ cells from their simulation.
   We find that ${\sim}4$ simulation cells per X-IFU pixel are enough
   to properly run our observational pipeline.
   It is possible that the number of cells extracted
   from the MHD simulation
   in a single direction of the sky plane
   is less than ${\sim}2$ per X-IFU pixel,
   depending on the dimensional injection length chosen to set the size of the simulation box.
   In such cases (in which both OBScustom and OBSfid lie),
   the extracted fluctuation fields need to be interpolated
   on a new grid with $128$ cells in each direction, so that
   ${\sim}4$ cells lie in the field of view of one X-IFU pixel
   as illustrated in Fig. \ref{fig:schematic}.
   This is why the final cells with volume
   $\ell_x\times\ell_y\times\ell_z$ are no longer cubic but
   rectangular with $\ell_y=\ell_z < \ell_x$.

   For both OBSfid and OBScustom, the length $L$ over which the cells are stacked
   in the line-of-sight direction
   (which is the physical length of the simulation box)
   may be smaller
   than the length scale over which significant emission takes place
   at $\x R_{200}$ in Perseus.
   We therefore rescale the integrated emissivity in the box in such a way that
   it matches the expected value from the integration of the squared density
   over the full line of sight, that is the total plasma emissivity (i.e. the emission measure)
   at $\x R_{200}$ in Perseus.
   This rescaling does not affect the root-mean square of the fluctuations
   after integration over the full line of sight themselves
   because the latter is self-consistently computed
   thanks to the toy-model Eq. (5) in \citet{clerc}
   (dotted lines in Fig. \ref{fig:3Drms}),
   and the amplitude of the fluctuation fields rescaled accordingly.
   
   At this stage, we stress that only the fluctuations of the thermodynamic fields,
   and not the global (background plus fluctuation) thermodynamic fields,
   are extracted from the MHD simulation and rescaled.
   The case of the velocity is different from the case of the thermodynamic fields
   since there is no bulk motion in the simulation
   and the field fluctuates around a zero average value.
   So we still need a proxy for the background temperature and density profiles
   by which the thermodynamic fluctuations will be weighted
   before being virtually observed.
   We chose the following local expression:
\begin{equation}
   X(z) = X_0\left(1-\frac{z}{H_X}\right),
\end{equation}
   with $X{=}T,\rho$
   and where $H_X$ is derived from the Perseus thermodynamic profiles
   presented in \ref{sec:prof}.
   With this choice, we are actually neglecting the cluster curvature
   by assuming a purely vertical dependency of
   the thermodynamic background profiles.
   With this hypothesis,
   the error made
   on the background
   temperature and density
   is less than $1 \%$
   when looking at $\x R_{200}$ of a cluster at
   Perseus redshift. However this error can grow up to $10\%$ for a similar density profile
   at $\zstroke_0 \gtrsim 0.08$.
   At higher redshift, the methodology later described in Appendix \ref{app:fluc}
   to isolate the fluctuations from the global fields
   can no longer be used as such, and should be modified to account for
   the cluster curvature.

   \subsection{Forward problem}
   \label{sec:forward}
   The purpose of the observation simulator SIXTE
   is essentially to create a photon list from a catalogue of point sources
   in the sky, according to the specifications of the instrument
   chosen for the virtual observation.
   A straightforward way to simulate the mock observation
   of an extended and resolved astrophysical source
   is to regard it as a collection of
   independent point sources
   More specifically, the initial spatial $(x,y,z)$-representation of the
   density, temperature and velocity fields
   need to be converted
   to a hybrid $(y,z,E)$-representation of the ideal spectra
   as a function of the sky position $(y,z)$, and
   where the energy variable $E$ ranges uniformly from $0.2$ to $12$ keV,
   covering the X-IFU energy range with a $1$-eV resolution.
   Each cell
   of this spatial representation
   is given a set of coordinates on the sky plane $(\vec{e_y},\vec{e_z})$
   according to the position of its centre in the mock cluster
   (there are 128 cells stacked in the line-of-sight direction $\vec{e_x}$,
   all with the same sky coordinates).
   Its X-ray line and Bremsstrahlung emission is modelled thanks to the APEC
   emission spectrum for collisionally-ionised diffuse gas \citep{smith}.
   After summing the spectra along the line of sight $\vec{e_x}$,
   we recover a hybrid $(y,z,E)$ representation by placing each
   integrated spectrum at its corresponding sky position.
   This catalogue of point sources is the input to the SIXTE simulator,
   which eventually leads to a mock observation of the ICM.
   In this process, the important instrumental and geometrical
   effects taken into account are the PSF, the response function,
   the detector geometry and its pixel filling-factor (${\sim} 97 \%$).
   Cross-talk and pile-up effects are simulated
   as well but marginal in our case,
   hence not accounted for in the post-processing step.
   More details about the SIXTE software and the response matrices
   of our baseline instrumental X-IFU configuration can respectively be found
   in \citet{dauser}\footnote{\url{https://www.sternwarte.uni-erlangen.de/sixte/}}
   and in \citet{barret16}\footnote{\url{http://x-ifu.irap.omp.eu/resources/for-the-community}}.
   Our setup does not simulate the vignetting.
   Therefore, the same response function can be used
   for all group of pixels in the post-processing step
   regardless of their off-axis position.
   
   The APEC model takes
   the metal abundance $Z$,
   the electron temperature $k_\mathrm{B} T$,
   the redshift $\zstroke$ and
   the norm $\mathcal{N}$
   as parameters.
   While the temperature is a direct output of the MHD simulation,
   the chemical abundance is not self-consistently computed and
   we simply assume a uniform metal abundance $Z=0.3Z_\odot$ for the modelling
   of the spectra, with $Z_\odot$ the solar value measured by \citet{anders}.
   The line-of-sight velocity field $\varv_{\mathrm{los}}=\varv_x$ is encoded in the redshift
   according to:
\begin{equation}
   \zstroke = (1+\zstroke_0)\sqrt{\frac{1+\frac{\varv_{x}}{c}}{1-\frac{\varv_{x}}{c}}} - 1,
\end{equation}
   with $c$ the speed of light in vacuum.
   The norm is expressed as:
\begin{equation}
   \mathcal{N} = 10^{-14} \frac{n_{\mathrm{e}} n_{\mathrm{H}} \mathcal{V}}{4\pi d_c^2(\zstroke)},
\end{equation}
   where $n_{\mathrm{e}}$ and $n_{\mathrm{H}}$,
   the electron and hydrogen number densities,
   are deduced from the density given the primordial abundances considered,
   $d_c(\zstroke)$ is the comoving distance
   and $\mathcal{V}{=}\ell_x\ell_y\ell_z$.

   \subsection{Inverse problem}
   \label{sec:inverse}
   Solving the inverse problem is to reverse-engineer the virtual observation
   in order to recover the thermodynamic and velocity fields.
   We recall that, for each observation, we divide the X-IFU grid
   into ${\sim}\Nbin$ Voronoi regions, from which a spectrum is extracted.
   These spectra are then fitted with the APEC model using the
   X-ray fitting package XSPEC.
   The spectral fit is done by minimising the Cash statistic \citep{cash}
   and without any a priori
   knowledge of the fitting parameters but the cluster
   cosmological redshift $\zstroke_0$.
   Since no vignetting effect is simulated in the forward problem,
   the same Ancillary Response File (ARF) is used for all
   Voronoi regions irrespective of their positions.
   At this step, the pixel filling-factor from the X-IFU
   must be taken into account in the ARF used for the fit in XSPEC,
   meaning that the response of all energy channels are $97\%$
   of the response from the ARF used by SIXTE to generate the mock observation.

   \section{Quantitative analysis of bias and dispersion}
   \label{app:analysis}

\begin{table*}
\caption{\label{t7}Bias and dispersion of
the output fields recovered from virtual observations with respect
to their input counterparts from the rescaled MHD simulation.}
\centering
\begin{tabular}{lcccccccc}
\hline\hline
Observation & $t_\mathrm{exp} \ (\mathrm{Ms})$ & $N_\mathrm{ph} \times 10^6$ & $b_{EM} \ (\%)$ & $b_{T_\mathrm{s}} \ (\%)$ & $b_{\varv_\mathrm{ew}} \ (\mathrm{km/s})$
 & $\sigma_{EM} \ (\%)$ & $\sigma_{T_\mathrm{s}} \ (\%)$ & $\sigma_{\varv_\mathrm{ew}} \ (\mathrm{km/s})$ \\
\hline
OBSfid    & 2 & 14.2 &-1.25 & -0.19 & -2.15
 & 2.32 & 1.64 & 19.2 \\
OBScustom & 2 & 14.9 & -1.39 & -0.57 & -12.0
 & 2.65 & 2.67 & 49.0 \\
\hline
\end{tabular}
\tablefoot{
          Column 1: name of virtual observation. 
          Column 2: exposure time.
          Column 3: total number of photons obtained.
          Column 4-6(7-9): average bias (dispersion) of the emission measure, spectroscopic temperature and emission-weighted velocity fields respectively.
          The values of the thermodynamic biases and dispersions are given
          in percents of their respective input quantities.
}
\label{tab:biasdisp}
\end{table*}

   The bias and standard deviation
   of the output fields coming from solving the inverse problem
   can be quantified,
   with respect to their input counterparts from the rescaled MHD simulation,
   by:
\begin{equation}
   b_{X} = <(X_\mathrm{obs} - X_\mathrm{sim})/X_\mathrm{sim}>,
   \label{eq:bias}
\end{equation}
\begin{equation}
   \sigma_{X}^2 = <\left(X_\mathrm{obs} - X_\mathrm{sim}\right)^2/X_\mathrm{sim}^2>,
   \label{eq:disp}
\end{equation}
   where $X$ is either the emission measure $EM$ or the spectroscopic temperature $T_\mathrm{s}$.
   In the case of the emission-weighted velocity $\varv_\mathrm{ew}$,
   the bias and dispersion are rather defined according to:
\begin{equation}
   b_{\varv_\mathrm{ew}} = <(\varv_\mathrm{ew,obs} - \varv_\mathrm{ew,sim})>,
   \label{eq:biasfluc}
\end{equation}
\begin{equation}
   \sigma_{\varv_\mathrm{ew}}^2 = <\left(\varv_\mathrm{ew,obs} - \varv_\mathrm{ew,sim}\right)^2>.
   \label{eq:dispfluc}
\end{equation}
   $<{\cdot}>$ is the spatial average running over all the Voronoi regions.
   The subscripts $\mathrm{obs}$
   and $\mathrm{sim}$ respectively stand for the observed and true input quantities.
   All these quantities
   are tabulated in Table \ref{tab:biasdisp}
   for the different synthetic observations.
   We do not necessarily seek to compare the quality of the reconstructed fields
   between the two different observations since they use different input physical fields anyway.
   We are rather aiming at validating the post-processing pipeline by checking that there is
   no significant bias or dispersion of the reconstructed fields
   with respect to the input quantities, independently for each observation.
   We are also looking for a quantitative criterion on the input fluctuation fields
   to predict whether they will be visually well recovered
   (as in OBScustom, Fig. \ref{fig:obscustom})
   or not (as in OBSfid, Fig. \ref{fig:obsfid}) with a X-IFU observation.

   The first row in Table \ref{tab:biasdisp} shows
   the biases and standard deviations
   of the reconstructed fields with respect
   to the input quantities in the case of OBSfid:
   both the thermodynamic
   and velocity fields are very well reconstructed
   with no significant biases and acceptable dispersions.
   More quantitatively, we find dispersions
   $\sigma_{T_\mathrm{s}}{\sim}2\%,\ \sigma_{\varv_\mathrm{ew}}{\sim}20$ km/s
   for the observed temperature and velocity field.
   This proves the robustness of our observational post-processing chain.
   Yet the fluctuation fields reconstructed from this synthetic observation
   are not necessarily visually satisfactorily retrieved,
   as seen in Fig. \ref{fig:obsfid}.
   The root-mean square of the input temperature fluctuation and velocity fields
   after integration along the line of sight
   $\left.\delta T_\mathrm{s}\right|_\mathrm{rms}{\sim}0.5\%,
   \ \left.\varv_\mathrm{ew}\right|_\mathrm{rms}{\sim}10\mathrm{km/s}$
   are smaller than the respective standard deviations
   $\sigma_{T_\mathrm{s}}{\sim}2\%,\ \sigma_{\varv_\mathrm{ew}}{\sim}20$ km/s
   of the reconstructed temperature and velocity fields with respect
   to the input quantities.
   We deduce that having $\left.\delta X\right|_\mathrm{rms}\gtrsim\sigma_X$, with $X= EM, T_\mathrm{s}$ or $\left.\varv_\mathrm{ew}\right|_\mathrm{rms}\gtrsim\sigma_{\varv_\mathrm{ew}}$ are necessary (albeit maybe not sufficient) conditions
   to satisfactorily detect the fluctuation fields from a synthetic observation.
   In this case, $\sigma_X$ can therefore be seen as a detectability threshold
   due to the inherent noise associated with a X-IFU observation:
   the fluctuation fields will be detected
   (i.e. visually well recovered)
   if their intensities
   exceed the noise level associated with the observation. 
   This is why,
   in the phenomenological discussion
   of Section \ref{sec:char}, Fig. \ref{fig:3Drms},
   we used
   $\sigma_{T_\mathrm{s}}{\sim}2\%$ and
   $\sigma_{\varv_\mathrm{ew}}{\sim}20$ km/s
   as the typical X-IFU detectability limits
   for a 2-Ms observation of the Perseus cluster at $\x R_{200}$.
   The case of the emission measure needs more caution
   because this extensive physical quantity is proportional to the squared density:
   when the root-mean square of the emission-measure and spectroscopic temperature fluctuations
   are small enough,
   they relate according to
   $\left. \delta EM \right|_\mathrm{rms}{\sim} 2\left. \delta T_\mathrm{s} \right|_\mathrm{rms}$.
   In the case of the emission measure fluctuation field, the factor two
   puts the root-mean square
   $\left. \delta EM \right|_\mathrm{rms}{\sim}1\%$
   closer to the corresponding detectability limit $\sigma_{EM}{\sim}2\%$.
   This explains why
   the emission measure fluctuations are, to some extent, visually better recovered
   than those of spectroscopic temperature
   in the case of the mock observation OBSfid
   (bottom maps in Fig. \ref{fig:obsfid}).

   The biases and dispersions of the velocity and thermodynamic fields
   reconstructed from the observation OBScustom
   are given in the second row of Table \ref{tab:biasdisp}.
   In the case of the thermodynamic fields,
   they are very similar to those found for the previous mock observation
   and prove once again the robustness of the reconstruction.
   However, the output velocity field
   is slightly biased towards negative values and more dispersed
   than in the first virtual observation.
   This is a consequence of the
   root-mean square of the input velocity field being
   larger in this synthetic observation:
   a similar behaviour was seen when the Mach number from an input hydrodynamic simulation
   was increased in
   \citet{roncarelli}.

\section{Computation of the thermodynamic fluctuations in the Voronoi regions}
\label{app:fluc}

   In this respect,
   we first derive the average thermodynamic quantity $\overline{X}$
   at fixed $z_i$
   (which represents an iso-gravity line when the cluster curvature is neglected,
   see Appendix \ref{app:design}):

\begin{equation}
   \overline{X}(z_i) = \frac{1}{n_y \left(z_i \right)} \sum_j X(y_j, z_i),
   \label{eq:averageX}
\end{equation}
   where
   $X$ is either
   the spectroscopic temperature $T_\mathrm{s}$
   or the emission measure $EM$.
   $j$ runs on all the X-IFU pixels of the $y$-axis
   at a given vertical position $z_i$ on the 2D grid
   and $n_{y}$ is the number
   of such pixels
   (which depends on $z_i$ because of X-IFU's hexagonal shape).
   The same Voronoi region can therefore
   be accounted for more than once in the average of a single row.
   The next step is to compute the average quantity $<X>$ in a single Voronoi region:
\begin{equation}
   <X> = \frac{1}{n_\mathrm{pix}} \sum_i \overline{X}(z_i),
   \label{eq:averageV}
\end{equation}
   where $n_\mathrm{pix}$ is the number of pixels in the Voronoi regions considered
   and $i$ runs on all those pixels, whose vertical position is $z_i$.
   We finally obtain the thermodynamic fluctuation $\delta X$
   in a given Voronoi region according to:
\begin{equation}
   \delta X = \frac{X - <X>}{<X>}.
   \label{eq:flucV}
\end{equation}
    We emphasise that this procedure
    is greatly simplified by the fact that the cluster curvature can be neglected.
    Ideally, it should return the same fluctuation field
    that we would get if we were to directly integrate the thermodynamic
    fluctuations themselves along the line of sight
    (not weighted by the background quantities, as done for the velocity field actually).
    We checked whether this procedure is robust or not in this sense for diverse fluctuation fields
    that we parametrise with different intensities and injection lengths.
    In the parameter ranges relevant to the present study,
    we find very little influence of these diverse parameters on the robustness
    of the method, which proves to be overall satisfying. 

\end{appendix}
\end{document}